\documentclass[10pt,twocolumn,twoside]{IEEEtran}
\usepackage{algorithm}
\usepackage{algpseudocode}

\usepackage{epsfig,latexsym}
\usepackage{float}
\usepackage{indentfirst}
\usepackage{amsmath}
\allowdisplaybreaks[4]
\usepackage{amssymb}
\usepackage{times}
\usepackage{subfigure}
\usepackage{pifont}
\usepackage{psfrag}
\usepackage{cite}
\usepackage{url}
\usepackage{lastpage}
\usepackage{bm}
\usepackage{color}
\usepackage{fancyhdr}
\usepackage[center]{caption2}
\usepackage{balance}
\usepackage[margin=0.75in]{geometry}
\newtheorem{theorem}{Theorem}

\newtheorem{Proposition}{Proposition}

\newtheorem{Lemma}{Lemma}

\ifCLASSINFOpdf  
\else
\fi

\begin{document}
	%
	\title{Energy Efficiency Optimization for Secure Transmission in MISO Cognitive Radio Network with Energy Harvesting}

	\author{Miao Zhang,~\IEEEmembership{Student Member,~IEEE,} Kanapathippillai Cumanan,~\IEEEmembership{Senior Member,~IEEE,} Jeyarajan Thiyagalingam,~\IEEEmembership{Member,~IEEE,} Wei Wang,~\IEEEmembership{Member,~IEEE,} Alister G. Burr,~\IEEEmembership{Senior Member,~IEEE,} Zhiguo Ding,~\IEEEmembership{Senior Member,~IEEE,} and Octavia A. Dobre,~\IEEEmembership{Senior Member,~IEEE}
		\thanks{This paper has been presented in part at the IEEE Wireless Communications and Signal Processing (WCSP), October 11-13, 2017 \cite{zhang2017secure}.} 
		\thanks{M. Zhang is with the School of Information Science and Technology, Nantong University, Nantong, China, also with the Department of Electronic Engineering, University of York, York, UK (email: mz1022@york.ac.uk).}
		\thanks{K. Cumanan and A. G. Burr are with the Department of Electronic Engineering, University of York, York, UK (email: \{ kanapathippillai.cumanan, alister.burr\}@york.ac.uk).}
		\thanks{J. Thiyagalingam is with the Scientific Computing Department of Rutherford Appleton Laboratory, Science and Technology	Facilities Council, Harwell Campus, Didcot, UK (email: t.jeyan@stfc.ac.uk).}
		\thanks{W. Wang is with the School of Information Science and Technology, Nantong University, Nantong, China, and with Research Center of Networks and Communications, Peng Cheng Laboratory, Shenzhen, China, and also with the Department of Electronic Engineering, University of York, York, United Kingdom (e-mail: wwang2011@ntu.edu.cn).}
		\thanks{Z. Ding is with the School of Electrical and Electronic Engineering, The University of Manchester, Manchester, UK (email: zhiguo.ding@manchester.ac.uk).}
		\thanks{O. A. Dobre is with the Department of Electrical and Computer Engineering, Memorial University, St. John’s, NL A1B 3X5, Canada (email: odobre@mun.ca).}
		\thanks{Corresponding author: Wei Wang (e-mail: wwang2011@ntu.edu.cn).}
	    \thanks{The work of M. Zhang, K. Cumanan, and A. G. Burr was supported by H2020-MSCA-RISE-2015 under grant number 690750. The work of W. Wang was was supported in part by the Stereoscopic Coverage Communication Network Verification Platform for China Sea under grant PCL2018KP002, the Six Categories Talent Peak of Jiangsu Province under grant KTHY-039, the Science and Technology Program of Nantong under grant GY22017013 and the Open Research Fund of Nantong University-Nantong Joint Research Center for Intelligent Information Technology under grant KFKT2017B02. The work of Z. Ding was supported by the UK EPSRC under grant number EP/N005597/2 and by H2020-MSCA-RISE-2015 under grant number 690750.}}

	
	%


	\maketitle
	
	\begin{abstract}
		In this paper, we investigate different secrecy energy efficiency (SEE)
		optimization problems in a multiple-input single-output underlay cognitive radio (CR) network in the presence of an energy harvesting receiver. In particular, these energy  efficient designs are developed with different assumptions of channels state information (CSI) at the transmitter, namely perfect CSI, statistical CSI and imperfect CSI with bounded channel uncertainties. 
		In particular, the overarching objective here is to design a beamforming technique maximizing the SEE while satisfying all relevant constraints linked to interference and harvested energy between transmitters and receivers. We show that the original problems are non-convex and their solutions are intractable. By using a number of  techniques, such as non-linear fractional programming and difference of concave (DC) functions, we  reformulate the original problems so as to render them tractable. We then combine these techniques with the Dinkelbach's algorithm to derive iterative algorithms to determine relevant beamforming vectors which lead to the SEE maximization. In doing this, we investigate the robust design with ellipsoidal bounded channel uncertainties, by mapping the original problem into a sequence of semidefinite programs by employing the semidefinite relaxation, non-linear fractional programming and S-procedure. Furthermore, we show that the maximum SEE can be achieved through a search algorithm in the single dimensional space. Numerical results, when compared with those obtained with existing techniques in the literature,  show the effectiveness of the proposed designs for SEE maximization. 
	\end{abstract}
	
	\IEEEpeerreviewmaketitle

	%

	\section{Introduction}
	\label{sec:intro}
	
	Wireless communications is one of the underpinning technologies of the
	modern society, and in fact it is not an overstatement to say that it
	plays an indispensable role in nearly all communication
	infrastructures. The fundamental requirements of future wireless
	networks include unprecedented higher data rates, ultra reliability
	and low latency along with the support for massive connectivity for
	handling the proliferation of Internet-of-Things. However, with
	the unprecedented growth in mobile data traffic in recent years, the
	overall demand for higher capacity and lower latency has increased
	rather tremendously~\cite{cirik2016beamforming,amin2015energy}. This, in turn, brings
	different challenges on the aspects surrounding the energy consumption
	in wireless networks. The energy consumption, regardless of the
	underpinning technologies, has a knock-on effect on the environment we
	live in, carbon footprint, global warming, and thus unanticipated
	financial consequences~\cite{hu2016optimal,al2019energy,bashar2019energy}.
	
	For this reason, the energy efficiency has become one of the crucial
	metrics for assessing the efficacy of wireless communication systems. As such, 
	optimizing the energy efficiency focusing on a particular
	type of wireless systems (such as multiple-input, multiple-output or multiple-input (MIMO), single-output (MISO)), is rather a common approach. A brief survey on both energy and spectrum efficiency is provided in \cite{hu2014energy}. The uplink energy efficiency for device-to-device multimedia cellular networks is investigated in \cite{wu2014energy}, whereas the authors in \cite{zhou2013energy} propose an energy-spectrum-aware scheduling scheme to study the tradeoff between energy efficiency and spectrum efficiency for mobile ad-hoc networks. The power allocation design for maximizing the energy efficiency in a cognitive radio network is considered in \cite{zhou2015energy}.
	
In this paper, we focus on optimizing the energy efficiency of an
underlay cognitive radio (CR) network for secure data transmission. In
particular, we pay special attention to energy efficiency within the
context of MISO systems with energy harvesting (EH)
requirement. Wireless EH is an emerging technology, which
  facilitates the mobile devices to collect energy from
  external energy sources without any wired connections
  \cite{varshney,grover2010shannon,ng2019wireless}. In general,
conventional EH methods harvest energy from natural sources, such as
wind, solar and
waves~\cite{hossain2014evolution,raghunathan2006emerging}, such an
extended approach is impractical for at least two reasons. First,
these external energy sources are unreliable and heavily correlated to
the environments conditions. Secondly, practically it is infeasible to
exploit them to the mobile devices, owing to the size limitations of
harvesting devices and the unstable power output caused by different
geographical conditions. As such, wireless EH facilitates practical
design and implementation especially in mobile
devices~\cite{wang2018multi,zhang2017secure, zhou2018enhancing,
  zhou2018resource,al2019on}. There exist three different schemes
  of technologies to implement wireless EH, name, magnetic
  induction, inductive coupling, and radio frequency-based wireless
  power transfer \cite{ng2019wireless, krikidis2014simultaneous}.  The
  first two schemes are primarily based on the near-field
  electromagnetic induction, and hence, do not have the capability to
  support long-range wireless power transfers \cite{ng2019wireless,
    krikidis2014simultaneous}. Thus, this paper focuses on the last
  scheme, which harvests energy from radio frequency signals that
  carry information. In \cite{li2019robust}, a robust energy
  efficiency design for a relay embedded MIMO system with SWIPT is
  presented. The secure beamforming design for a cooperative MISO
  cognitive radio non-orthogonal multiple access network with SWIPT
  has been investigated in \cite{zhou2018artificial}. Our focus on
this paper is within the scope of wireless energy harvesting in a MISO
CR network.
	
	The other aspect in which the contributions of this manuscript are linked is information security. This represents one of the major issues in wireless networks, as the signals transmitted through the wireless medium are more vulnerable
	for interception. Most commonly used conventional security methods
	completely rely on classic cryptographic techniques implemented at upper
	layers. The fundamental concept of these methods is to employ a set of
	secret keys to encrypt plain-text at the transmission end, and to decrypt
	ciphertext at the receiving end. Although existing traditional
	security techniques may remain non-contestable, the broadcast nature of wireless
	communications introduces a number of challenges, particularly in terms of key exchange and
	distributions~\cite{cumanan2014secrecy,zhu2018secure, zhang2016secrecy,cumanan2016secrecy, cumanan2016physical,cumanan2017secure,cumanan2012new,alavi2019robust,alavi2017limited,xiang2015signal,cumanan2009sinr,cumanan20019sinr}. To further enhance the security of wireless networks,  information theoretic-based physical
	layer security has been proposed to complement the conventional
	security techniques in wireless
	transmissions. Physical layer security can  only be applied in an
	environment where the signal-to-noise ratio (SNR) of the legitimate
	channel is better than that of the eavesdropper
	channel~\cite{Shannon,Wyner,csiszar}. In particular, this approach
	exploits the nature of physical layer dynamics in establishing secure
	wireless transmission~\cite{wang2016achieving}. In contrast to the
	conventional security techniques, the application of physical layer
	security schemes makes it more difficult for interceptors to eavesdrop
	and decode the information  intended for legitimate nodes. The beamforming design for physical layer security over $\eta-\mu$ fading channels is considered with two different scenarios, namely with and without co-channel interference in \cite{yang2018physical}. In \cite{li2019beamforming}, the sum secrecy rate maximization problem is studied for a relay-based cognitive radio network.
	
	Most of the literature on physical layer security focuses primarily on efficiently utilizing the required transmit power to offer different quality-of-service requirements, such as  achieving better secrecy rates~\cite{zhang2018robust,cumanan2014secrecy,chu2016simultaneous,chu2016secrecy,chu2015robust,chu2017robust,nguyen2016joint,cumanan2010multiuser,cumanan2010joint,cumanan2009sinr,}. However, these designs do not consider the secrecy energy efficiency (SEE), which is a suitable performance metric for measuring the efficient utilization of the power consumption in secure communication systems. The SEE is defined as the ratio between the achieved secrecy rate and the total power consumption at the transmitter. A few works exist on SEE~\cite{mei2016robust,mei2017energy,mei2017outage}. SEE maximization with statistical eavesdroppers’ channel state information (CSI) has been considered in~\cite{mei2017energy,mei2017outage}, whereas SEE maximization with ellipsoidal based channel uncertainties has been investigated in~\cite{mei2016robust}. However, none of these works consider EH in a CR network.
	
	Motivated by the absence of a combined approach, in this paper, we solve a number of  SEE maximization problems for an underlay MISO CR network with EH requirement. In particular, a multi-antenna secondary transmitter (SU-Tx) simultaneously sends confidential information and energy to a secondary receiver (SU-Rx) and an energy receiver (ER), respectively. This secondary simultaneous wireless information and power transfer communication is established by sharing the spectrum that is allocated for communication for primary user terminals, as shown in Figure  1. In particular, we consider transmit beamforming design to maximize the achievable SEE under the constraints of secrecy rate on the SU-Rx, interference leakage on the primary receiver (PU-Rx) and EH requirement on ER. Furthermore, the ER is considered to be a potential eavesdropper due to the broadcast nature of wireless transmission.

	\noindent The key contributions of this work are summarized as follows:
	
	\begin{enumerate}
		
		\item With the assumption of perfect CSI at the SU-Tx, we formulate the transmit beamforming design as an SEE maximization problem with the required set of constraints. The original SEE maximization problem is not convex in nature due to its non-linear fractional objective function. To circumvent this non-convex issue, we transform the original problem into a tractable form by exploiting different mathematical techniques including non-linear fractional programming~\cite{dinkelbach1967nonlinear} and difference of concave (DC) functions programming~\cite{dinh2014recent}. Furthermore, we propose an iterative algorithm to yield an optimal solution and the optimality of the solution is proven.
		
		\item Next, we study the SEE maximization with the statistical CSI available at the SU-Tx. In particular, we design the transmit beamforming vectors to maximize the SEE, while satisfying the constraints on interference leakage, outage probability on secrecy rate and EH requirements. This problem appears to be challenging to solve due to the probabilistic constraints, which are difficult to mathematically define in the design. We express these outage constraints into a set of closed-form expressions. Then, the original problem is efficiently solved through an iterative approach by exploiting non-linear fractional programming and DC programming. Furthermore, we also prove that the optimal solution obtained by our proposed method always yields rank-one and shows a similar performance of the semidefinite relaxation (SDR) approach. 
		
		\item Finally, we consider a robust design with ellipsoidal based channel uncertainties for all channels. This robust SEE maximization problem is non-convex in its original form and we first reformulate it into a series of semidefinite programs (SDP) by employing the SDR and non-linear fractional programming~\cite{dinkelbach1967nonlinear}. However, the reformulated problem still remains non-convex due to the channels uncertainties. In order to overcome this non-convexity issue, we exploit the S-procedure \cite{boyd1994linear} to convert this problem into a convex one~\cite{boyd1994linear}. Moreover, we also provide a method to construct a rank-one optimal covariance matrix. This confirms the optimality of the proposed robust SEE maximization based beamforming design.
		
	\end{enumerate}

       The remainder of this paper is organized as follows. The system model is presented in Section II. The SEE maximization problem with perfect CSI assumption is formulated and iterative algorithms are proposed to solve it in Section III. The robust SEE maximization problems with the statistical and imperfect CSI assumptions are formulated and solved in Sections IV and V, respectively. Simulation results for the performance of the proposed algorithms are provided in Section VI. Finally, concluding remarks are drawn in Section VII.

	\subsection{Notations}
	
	We use the upper and lower case boldface letters for matrices and vectors, respectively. $(\cdot)^{-1}$, $(\cdot)^T$ and $(\cdot)^H$ stand for inverse, transpose and conjugate transpose operation, respectively. $\mathbf{A}\succeq\mathbf{0}$ means that $\mathbf{A}$ is a positive semidefinite matrix. $\textrm{rank}(\mathbf{A})$ denotes the rank of a matrix, and $\textrm{tr}(\mathbf{A})$ represents the trace of matrix $\mathbf{A}$. The circularly symmetric complex Gaussian distribution is represented by $\mathcal{CN}(\mu,\sigma^2)$ with mean $\mu$ and variance $\sigma^2$. $\mathbb{H}^{N}$ denotes the set of all $N \times N$ Hermitian matrices. $\textrm{ln}(x)$ is the natural logarithm of $x$.

	\section{System Model}
	
\begin{center}
		\begin{figure}[t!]
			\centering\includegraphics[width=\linewidth]{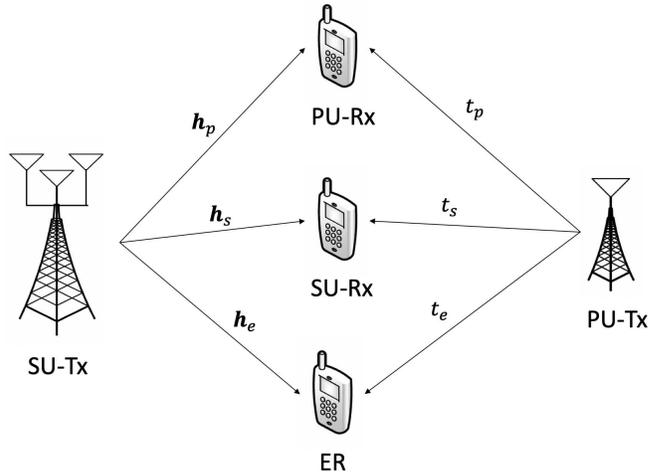}
			\caption{An underlay CR network with a multi-antenna SU-Tx and single antenna PU-Tx, PU-Rx, SU-Rx and ER. }	
			\label{fig:SRP}
		\end{figure}
	\end{center}

	Figure~\ref{fig:SRP} shows the system model, which we use as vehicle for our contributions in this work. We consider a downlink transmission of MISO CR network with five terminals: an SU-Tx, an SU-Rx, a primary transmitter (PU-Tx), a PU-Rx and an ER. The SU-Tx is equipped with $N_{t}$ antennas, while the ER,  SU-Rx,  PU-Tx and PU-Rx have a single antenna, respectively. 
	\vspace{-2em}
	
	We assume that the multi-antenna SU-Tx intends to send confidential information to the single antenna SU-Rx while satisfying the interference leakage threshold on the PU-Rx. The ER harvests energy from the SU-Tx transmission through wireless power transfer technique. However, the ER might as well be a potential eavesdropper and may attempt to intercept the message intended to the SU-Rx. Therefore, the ER is assumed to be a potential eavesdropper in this CR network. The channel coefficients between the SU-Tx and PU-Rx, SU-Rx and ER are denoted by $\mathbf{h}_{p}\in\mathcal{C}^{N_{T}\times1}$, $\mathbf{h}_{s}\in\mathcal{C}^{N_{T}\times1}$ and $\mathbf{h}_{e}\in\mathcal{C}^{N_{T}\times1}$, respectively. Thus, the received signal at SU-Rx and ER can be expressed as
	\begin{align}
	y_{s}=\mathbf{h}_{s}^{H}\mathbf{q}+t_{s}\sqrt{\Psi}x+n_{s},\\
	{y}_{e}=\mathbf{h}_{e}^{H}\mathbf{q}+t_{e}\sqrt{\Psi}x+{n}_{e},
	\end{align}
	\noindent where $\mathbf{q}\in\mathcal{C}^{N_{T}\times1}$ denotes the beamforming vector from the SU-Tx. The notations $n_{s}\sim\mathcal{CN}(0,\sigma_{s}^{2})$ and ${n}_{e}\sim\mathcal{CN}(0,\sigma_{e}^{2})$ represent the joint effects of
	thermal noise and signal processing noise, at the SU-Rx and the ER, respectively. Moreover, we assume that $\Psi$ and $x\in\mathcal{C}^{1\times1}$ are the transmit power of PU-Tx and the information signal intended to the PU-Rx, respectively. The $t_{s}\in\mathcal{C}^{1\times1}$ and $t_{e}\in\mathcal{C}^{1\times1}$ denote the channel coefficients between the PU-Tx and SU-Rx as well as ER, respectively. Moreover, the equivalent noise at the SU-Rx and ER, which capture the joint effect of the received interference from the PU-Tx, can be modelled as additive white Gaussian noise with zero mean and $\sigma_{su}^{2}$ and $\sigma_{er}^{2}$ variances, respectively \cite{ng2016multiobjective,pei2010secure,bedeer2014multiobjective}. The achievable secrecy rate at the SU-Rx is defined as
	\begin{align}
	R_{s}\!=\!\log_{2}(1+\frac{1}{\sigma_{su}^{2}}\mathbf{h}_{s}^{H}\mathbf{q}\mathbf{q}^{H}\mathbf{h}_{s})-\log_{2}(1+\frac{1}{\sigma_{er}^{2}}\mathbf{h}_{e}^{H}\mathbf{q}\mathbf{q}^{H}\mathbf{h}_{e}).
	\end{align}
	The total transmit power consumption at the SU-Tx is
	\begin{align}
	P_{t}=\frac{||\mathbf{q}||^{2}+P_{c}}{\xi},
	\end{align}
	where $P_{c}$ is the circuit power consumption of the transmitter and $\xi\in(0,1]$ is the power amplifier efficiency, which is assumed to be one  ($\xi=1$) throughout this paper without loss of generality. The SEE, which is defined as the ratio between the achievable secrecy rate and the total transmit power consumption, can be expressed as
	\begin{align}
	\eta&=\frac{R_{s}}{P_{t}}\nonumber\\
	&=\frac{\log_{2}(1+\frac{1}{\sigma_{su}^{2}}\mathbf{h}_{s}^{H}\mathbf{q}\mathbf{q}^{H}\mathbf{h}_{s})-\log_{2}(1+\frac{1}{\sigma_{er}^{2}}\mathbf{h}_{e}^{H}\mathbf{q}\mathbf{q}^{H}\mathbf{h}_{e})}{||\mathbf{q}||^{2}+P_{c}}.
	\end{align}
	The interference leakage to the PU-Rx is defined as
	\begin{align}
	P_{il}=\mathbf{h}_{p}^{H}\mathbf{q}\mathbf{q}^{H}\mathbf{h}_{p}.
	\end{align}
Recently, non-linear EH models have been investigated in the literature \cite{boshkovska2015practical, boshkovska2016robust, ni2018outage}, which are more realistic and practical models to reflect the non-linear relationship between the radio frequency and direct current conversion circuits. However, a generic non-linear EH model that could address all the issues in practical EH scenarios is still non-existent~\cite{zeng2017communications,wang2017joint}. Furthermore, by observing the measurement data presented in \cite{boshkovska2017robust, stoopman2013self}, the output power of the linear EH model is accurate when compared against the non-linear model, even when considered across wider ranges of the input radio frequency power. The linear EH model is also an effective model and has been widely adopted in the analysis and optimal design for EH systems  \cite{chu2016simultaneous,chu20177robust,zhu2016joint,zhu2018secure, wang2017beamforming, wang2017optimal}.  By  assuming that the input radio frequency power is well within the linear regime of the rectifier, we  focus on a linear EH model. The
        harvested energy at the ER can be written as
	\begin{align}
	P_{r}=\zeta_{eh}\mathbf{h}_{e}^{H}\mathbf{q}\mathbf{q}^{H}\mathbf{h}_{e},
	\end{align}
	where $\zeta_{eh}\in(0,1]$ represents the EH efficiency of the ER. 
	
	\section{SEE Maximization with Perfect CSI}
	\label{sec:see-max-pcsi}
	
	In this section, we solve an SEE maximization problem with the minimum harvested energy and the maximum interference leakage constraints at the ER and PU-Rx, respectively. It is assumed that perfect CSI of all terminals is available at the SU-Tx.  Note that for the passive ER, the CSI can be estimated by the SU-Tx through local oscillator power leakage from the ER's RF front end, which is described in detail in \cite{mukherjee2012detecting}. This SEE maximization problem is formulated as
	\begin{subequations}\label{eq:SEE_max_ori}
		\begin{align}
		\max_{\mathbf{Q}_{s}}&~ \frac{\log_{2}(1\!+\!\frac{1}{\sigma_{su}^{2}}\mathbf{h}_{s}^{H}\mathbf{q}\mathbf{q}^{H}\mathbf{h}_{s})\!-\!\log_{2}(1\!+\!\frac{1}{\sigma_{er}^{2}}\mathbf{h}_{e}^{H}\mathbf{q}\mathbf{q}^{H}\mathbf{h}_{e})}{||\mathbf{q}||^{2}+P_{c}} \label{eq:SEE_max_obj} \\
		s.t.& \log_{2}(1+\frac{1}{\sigma_{su}^{2}}\mathbf{h}_{s}^{H}\mathbf{q}\mathbf{q}^{H}\mathbf{h}_{s})-\log_{2}(1+\frac{1}{\sigma_{er}^{2}}\mathbf{h}_{e}^{H}\mathbf{q}\mathbf{q}^{H}\mathbf{h}_{e})\nonumber\\
		& \geq R_{d}, \label{eq:SEE_Max_c1}\\
		&\zeta_{eh}\mathbf{h}_{e}^{H}\mathbf{q}\mathbf{q}^{H}\mathbf{h}_{e} \geq \omega_{s}, \label{eq:SEE_Max_c2}\\
		& \mathbf{h}_{p}^{H}\mathbf{q}\mathbf{q}^{H}\mathbf{h}_{p}\leq P_{f},  ||\mathbf{q}||^{2} \leq P_{\textrm{tx}},  \label{eq:SEE_Max_c3}
		\end{align}
	\end{subequations}
	\noindent where $\omega_{s}$ indicates the minimum EH requirement at the ER. $P_{f}$ and $P_{\textrm{tx}}$ are the predefined interference leakage tolerance at the PU-Rx and the maximum available transmit power at the SU=Tx, respectively.
	Let  $\mathbf{Q}_{s}$ be a rank-one covariance matrix,  based on the beamforming vector $\mathbf{q}$ such that $\mathbf{Q}_{s}=\mathbf{q}\mathbf{q}^{H}$. Then the original problem can be expressed as 
	\begin{subequations}\label{eq:SEE_max_relax}
		\begin{align}
		&\max_{\mathbf{Q}_{s}}~ \frac{R_{s}(\mathbf{Q}_{s})}{P_{t}(\mathbf{Q}_{s})} \label{eq:SEE_max_reobj} \\
		s.t.&~ \log_{2}(1\!+\!\frac{1}{\sigma_{su}^{2}}\mathbf{h}_{s}^{H}\mathbf{Q}_{s}\mathbf{h}_{s})\!-\!\log_{2}(1\!+\!\frac{1}{\sigma_{er}^{2}}\mathbf{h}_{e}^{H}\mathbf{Q}_{s}\mathbf{h}_{e})\geq R_{d}, \label{eq:SEE_Max_rec1}\\
		&~\zeta_{eh}\mathbf{h}_{e}^{H}\mathbf{Q}_{s}\mathbf{h}_{e} \geq \omega_{s}, \label{eq:SEE_Max_rec2}\\
		&~ \mathbf{h}_{p}^{H}\mathbf{Q}_{s}\mathbf{h}_{p}\leq P_{f}, \label{eq:SEE_Max_rec4}\\
		&~ \textrm{tr}(\mathbf{Q}_{s}) \leq P_{\textrm{tx}}, \mathbf{Q}_{s}\succeq \mathbf{0}, \textrm{rank}(\mathbf{Q}_{s})=1,  \label{eq:SEE_Max_rec3}
		\end{align}
	\end{subequations}
	
	\noindent where $$
	R_{s}(\mathbf{Q}_{s})=\log_{2}(1\!+\!\frac{1}{\sigma_{su}^{2}}\mathbf{h}_{s}^{H}\mathbf{Q}_{s}\mathbf{h}_{s})\!-\!\log_{2}(1\!+\!\frac{1}{\sigma_{er}^{2}}\mathbf{h}_{e}^{H}\mathbf{Q}_{s}\mathbf{h}_{e}) 
	$$ and
	$P_{t}(\mathbf{Q}_{s})=\textrm{tr}(\mathbf{Q}_{s})+P_{c}.$
	
	This is a non-convex problem due to the fractional objective function. To recast this problem as a convex one, we exploit non-linear fractional and DC programming in the following subsections.
	
	\subsection{Non-linear Fractional Programming}
	
	The objective function defined in (\ref{eq:SEE_max_reobj}) is a fractional function with non-linear terms both in the numerator and denominator. Furthermore, the overall problem presented in (\ref{eq:SEE_max_relax}) is known as a non-linear fractional problem in the literature~\cite{dinkelbach1967nonlinear}. First, we transform it into a parametric programming. 
	
	\begin{theorem}\label{proof:nonlinear}
		Let $R_{s}(\mathbf{Q}_{s}^{*})>0$ and $P_{t}(\mathbf{Q}_{s}^{*})>0$ be the optimal achieved secrecy rate and optimal minimum total power consumption of the problem defined in (\ref{eq:SEE_max_relax}), respectively. Further, let $\mathcal{F}$ be the feasible solution set of this problem. Moreover, the transmit covariance matrix $\mathbf{Q}^{*}_{s}$ achieves the maximum energy efficiency such that
		\begin{align}
		\lambda^{*}=\frac{R_{s}(\mathbf{Q}^{*}_{s})}{P_{t}(\mathbf{Q}^{*}_{s})}=\max_{\mathbf{Q}_{s}\in\mathcal{F}} \bigg\{\frac{R_{s}(\mathbf{Q}_{s})}{P_{t}(\mathbf{Q}_{s})}\bigg\}, 
		\end{align}
		if and only if $R_{s}(\mathbf{Q}^{*}_{s})$, $P_{t}(\mathbf{Q}_{s}^{*})$ and $\lambda^{*}$ satisfy the following condition:
		\begin{align}
		\max_{\mathbf{Q}_{s}\in\mathcal{F}} \{R_{s}(\mathbf{Q}_{s})-\lambda^{*}P_{t}(\mathbf{Q}_{s})\}= R_{s}(\mathbf{Q}^{*}_{s})-\lambda^{*}P_{t}(\mathbf{Q}^{*}_{s})=0. \label{conditional1}
		\end{align}
		Additionally,
		\begin{align}
		~\max_{\mathbf{Q}_{s},\lambda} \{R_{s}(\mathbf{Q}_{s})-\lambda P_{t}(\mathbf{Q}_{s})\}\label{Theo1}
		\end{align}
		is defined as a parametric programming with parameter $\lambda$~\cite{dinkelbach1967nonlinear}.
	\end{theorem}
	\begin{IEEEproof}
		Please refer to Appendix \ref{proof_of_theorem}.
	\end{IEEEproof}
	~\\
	\begin{Lemma}
		The objective function defined in (\ref{Theo1}) is a convex, strictly decreasing and continuous function with respect to (w.r.t.) $\lambda$.
	\end{Lemma}
	\begin{Lemma}
		The equation $\max_{\mathbf{Q}_{s},\lambda} \{R_{s}(\mathbf{Q}_{s})-\lambda P_{t}(\mathbf{Q}_{s})\}=0$ has a unique solution, and let denote it by $\lambda_{0}$. Then, the problems defined in (\ref{Theo1}) and (\ref{eq:SEE_max_relax}) have the same optimal solution with the optimal SEE of $\lambda_{0}$.
	\end{Lemma}
	\begin{IEEEproof}
		Please refer to~\cite{dinkelbach1967nonlinear}.
	\end{IEEEproof}
	~\\
	By utilizing Theorem 1, we recast the original problem defined in (\ref{eq:SEE_max_relax}) into the following parametric programming problem with parameter $\lambda$:
	\begin{align}\label{nonlin}
	&F(\lambda)=\max_{\mathbf{Q}_{s}}[R_{s}(\mathbf{Q}_{s})-\lambda P_{t}(\mathbf{Q}_{s})]\nonumber\\
	&=\max_{\mathbf{Q}_{s}}\{\log_{2}(1+\frac{1}{\sigma_{su}^{2}}\mathbf{h}_{s}^{H}\mathbf{Q}_{s}\mathbf{h}_{s})-\log_{2}(1+\frac{1}{\sigma_{er}^{2}}\mathbf{h}_{e}^{H}\mathbf{Q}_{s}\mathbf{h}_{e})\nonumber\\
	&-\lambda[ \textrm{tr}(\mathbf{Q}_{s})+P_{c}]\}\nonumber\\
	&s.t. ~\textrm{(\ref{eq:SEE_Max_rec1})-(\ref{eq:SEE_Max_rec3})}. 
	\end{align}
	It can be seen that the original problem presented in (\ref{eq:SEE_max_relax}) is transformed into a parameterized polynomial subtractive form. As a result, the original problem is reformulated to determine $\lambda^{*}$ and $\mathbf{Q}_{s}^{*}$, which can satisfy the condition provided in (\ref{conditional1}) . Furthermore, by utilizing Dinkelbach's algorithm~\cite{dinkelbach1967nonlinear} with an initial value $\lambda_{0}$ of $\lambda$, the optimal solutions of  (\ref{eq:SEE_max_relax}) can be obtained by iteratively solving the following optimization problem:
	\begin{align}\label{nonlp}
	\max_{\mathbf{Q}_{s}}&~[R_{s}(\mathbf{Q}_{s})-\lambda_{i} P_{t}(\mathbf{Q}_{s})]\nonumber\\
	s.t.&~\textrm{(\ref{eq:SEE_Max_rec1})-(\ref{eq:SEE_Max_rec3})},
	\end{align}
	for a given $\lambda_{i}$ at the $i$th iteration. The value of $\lambda_{i}$ can be considered as the SEE obtained at the previous iteration. At each iteration, $\lambda_{i+1}$ should be updated such that 
	\begin{align}
	\lambda_{i+1}=\frac{R_{s}^{i}(\mathbf{Q}_{s})}{P_{t}^{i}(\mathbf{Q}_{s})},
	\end{align}
	where $R_{s}^{i}$ and $P_{t}^{i}$ denote the achieved secrecy rate and total power consumption of (\ref{nonlp}) for a given $\lambda_{i}$ from the previous iteration, respectively. This iterative process will be terminated when the condition in (\ref{conditional1}) is satisfied. However, in practice the iterative process will be carried out until the following inequality is satisfied:
	\begin{align} 
	\Delta F(\lambda)=|R_{s}^{i}(\mathbf{Q}_{s})-\lambda_{i}P_{t}^{i}(\mathbf{Q}_{s})|\leq \varepsilon,\label{convergence_V1}
	\end{align}
	\noindent where $\varepsilon>0$ is the convergence threshold.

	\subsection{DC Programming}
	
	In this subsection, we provide the required details on DC programming to solve the SEE maximization problem. DC programming is a well-known optimization approach to solve non-convex problems. In particular, this technique can be exploited in an optimization problem with an objective function defined as a difference of two concave functions. Since the objective function in (\ref{nonlp}) falls under this category, we utilize DC programming to solve this problem.
	
	The fundamental idea of DC programming is to locally linearize the non-concave functions at a feasible point $\mathbf{Q}_{s}^{k}$, and to iteratively update the approximation by solving the corresponding approximated problem~\cite{dinh2014recent}. We define the following function to approximate the second term of the objective function in (\ref{nonlp}):
	\begin{align}
	f(\mathbf{Q}_{s},\mathbf{Q}_{s}^{k})=&\log_{2}(1+\frac{1}{\sigma_{er}^{2}}\mathbf{h}_{e}^{H}\mathbf{Q}_{s}^{k}\mathbf{h}_{e})+\nonumber\\&\frac{\frac{1}{\sigma_{er}^{2}}\mathbf{h}_{e}^{H}(\mathbf{Q}_{s}-\mathbf{Q}_{s}^{k})\mathbf{h}_{e}}{(1+\frac{1}{\sigma_{er}^{2}}\mathbf{h}_{e}^{H}\mathbf{Q}_{s}^{k}\mathbf{h}_{e})\ln2}.
	\end{align}
	Based on this approximation, the problem defined in (\ref{nonlp}) can be transformed into an  equivalent problem as follows:
	\begin{align}
	\max_{\mathbf{Q}_{s}}&\{\log_{2}(1\!+\!\frac{1}{\sigma_{su}^{2}}\mathbf{h}_{s}^{H}\mathbf{Q}_{s}\mathbf{h}_{s})\!-\!f(\mathbf{Q}_{s},\mathbf{Q}_{s}^{k})\!-\!\lambda_{i}[ \textrm{tr}(\mathbf{Q}_{s})\!+\!P_{c}]\}\nonumber\\
	s.t. &~\textrm{(\ref{eq:SEE_Max_rec1})-(\ref{eq:SEE_Max_rec3})}.
	\end{align}
	This problem is still non-convex due to the non-convex rank-one constraint. By employing SDR, we relax the problem in (\ref{eq:SEE_max_relax}) by dropping the rank constraint $\textrm{rank}(\mathbf{Q}_{s})=1$, which can be defined as:
	\begin{align}
	\max_{\mathbf{Q}_{s}}&\{\log_{2}(1\!+\!\frac{1}{\sigma_{su}^{2}}\mathbf{h}_{s}^{H}\mathbf{Q}_{s}\mathbf{h}_{s})\!-\!f(\mathbf{Q}_{s},\mathbf{Q}_{s}^{k})\!-\!\lambda_{i}[ \textrm{tr}(\mathbf{Q}_{s})\!+\!P_{c}]\} \nonumber\\
	s.t.&~ \textrm{(\ref{eq:SEE_Max_rec1})-(\ref{eq:SEE_Max_rec4}}),~ \mathbf{Q}_{s}\succeq \mathbf{0},  \textrm{tr}(\mathbf{Q}_{s}) \leq P_{\textrm{tx}}.  \label{eq:SEE_max_relaxed}
	\end{align}
	\begin{Proposition}\label{proposition:rank_proof1}
		Provided that the problem (\ref{eq:SEE_max_relaxed}) is feasible, the optimal solution will be always rank-one.
	\end{Proposition}
	\begin{IEEEproof}
		Please refer to Appendix \ref{proof_of_proposition1}.
	\end{IEEEproof}
	~\\
	
	This approximated problem is convex in terms of $\mathbf{Q}_{s}$. Hence, the suboptimal solution $\mathbf{Q}_{s}^{*}$ to the original problem can be obtained by solving the problem in (19) and iteratively updating $\mathbf{Q}_{s}^{k}$ based on the solution obtained from the previous iteration. The algorithm based on DC programming is summarized in Algorithm 1.
	\begin{algorithm}[h]
		\caption{Iterative algorithm to solve (\ref{eq:SEE_max_relaxed})}
		\label{alg:DC}		
		\begin{algorithmic}[1]
			\State Initialize $i=0$ and choose an initial value $\lambda_{0}$;
			\Repeat $\leftarrow$ Outer loop (Dinkelbach's algorithm)
			\State Initial $k=0$, choose an initial value $\mathbf{Q}_{s}^{k}=0$ and \phantom a \phantom a \phantom a $F(\lambda)^{i,k}=0$;
			\Repeat $\leftarrow$ Inner loop (DC programming)
			\State Solve the problem (\ref{eq:SEE_max_relaxed}) with $\lambda=\lambda_{i}$ and obtain \phantom a \phantom a\phantom a \phantom a\phantom a \phantom a$\mathbf{Q}_{s}^{k+1}$;
			\State Compute  $F(\lambda)^{i,k+1}=\log_{2}(1+\frac{1}{\sigma_{su}^{2}}\mathbf{h}_{s}^{H}\mathbf{Q}_{s}^{k+1}\mathbf{h}_{s})$\phantom a \phantom a\phantom a \phantom a\phantom a \phantom . $-\log_{2}(1+\frac{1}{\sigma_{su}^{2}}\mathbf{h}_{e}^{H}\mathbf{Q}_{s}^{k+1}\mathbf{h}_{e})- \lambda_{i}[\textrm{tr}(\mathbf{Q}_{s}^{k+1})+P_{c}]$;
			\State $\Delta\mu=F(\lambda)^{i,k+1}-F(\lambda)^{i,k}$;
			\State Update $k=k+1$;
			\Until $|\Delta \mu|\leq \zeta$;
			\State Update $\lambda_{i+1}$ through (15);		
			\State $i=i+1$;
			\Until (\ref{convergence_V1}) satisfied;
			\State Return $\lambda^{*}=\lambda_{i}, P^{*}_{t}(\mathbf{Q}_{s})=P_{t}^{i-1}(\mathbf{Q}_{s}), R_{s}^{*}=R_{s}^{i-1}$;
			\State Output $\lambda_{i}$.
		\end{algorithmic}
	\end{algorithm}
\subsection{Computational Complexity}
	In this subsection, we provide the computational complexity analysis of Algorithm~1. The problem defined in (19) has four linear constraints, and one linear matrix inequality (LMI) constraint of size $N_{T}$. The number of variables $n$ is in the order of $N_{T}^{2}$. To obtain an optimal solution to the problem defined in (19), the computational complexity is in the order of $\sqrt{\varUpsilon(\Bbbk)}\Lambda\ln(1/\epsilon)$, where $\varUpsilon(\Bbbk)=N_{T}+4$, $\epsilon>0$, and $\Lambda=n^{3}+n(N_{T}^{3}+4)+n^{2}(N_{T}^{2}+4)$ \cite{wang2014outage, ben2001lectures}. The total computational complexity of solving the problem defined in (9) can be calculated through multiplying the computational cost of solving (19) by the number of both inner and outer loop iterations. Therefore, the computational complexity of solving the problem defined in (9) should be in the order of $TK\sqrt{N_{T}+4}[n^{3}+n(N_{T}^{3}+4)+n^{2}(N_{T}^{2}+4)]\ln(1/e),$ where $T$ and $K$ are the numbers of iterations of the inner and outer loop, respectively.
	
	\section{SEE Maximization with Statistical CSI}
	\label{sec:see_max_scsi}
	
	In this section, we consider a robust design with a realistic assumption that only the statistical CSI of ER is available at the SU-Tx. This scenario could arise when ER is part of neither the primary nor the secondary system. Hence, it is difficult to have ER's CSI at the SU-Tx based on handshaking signals. Furthermore, this assumption can be further supported by the fact that the ER might be silent or passive to hide its existence from the SU-Tx. Hence, we assume that only the statistical information on ER's CSI is available at the SU-Tx and the corresponding ER's CSI can be defined as 
	\begin{align}
	\mathbf{h}_{e}\sim\mathcal{CN}(\mathbf{0},\mathbf{G}_{e}),
	\end{align}
	where $\mathbf{G}_{e}\succeq \mathbf{0}$ is the covariance matrix of $\mathbf{h}_{e}$. With this statistical CSI, the SEE maximization problem can be formulated as follows:
	\begin{subequations}\label{SEE_max_st_ori}
		\begin{align}
		\max_{\mathbf{Q}_{s}} &~\delta \\
		s.t.&~\textrm{Pr}\{f_{s}(\mathbf{Q}_{s})\geq \delta\}\geq 1-\alpha, \label{SEE_max_st_c1}\\ 
		&~\textrm{Pr}\{g_{s}(\mathbf{Q}_{s})\geq R_{d}\}\geq 1-\beta,\label{SEE_max_st_c2}\\
		&~ \textrm{Pr}\{\zeta_{eh}\mathbf{h}_{e}^{H}\mathbf{Q}_{s}\mathbf{h}_{e} \leq \omega_{s}\} \leq \gamma, \label{SEE_max_st_c3}\\
		&~ \mathbf{h}_{p}^{H}\mathbf{Q}_{s}\mathbf{h}_{p}\leq P_{f},  \label{SEE_max_st_c4}\\
		&~\textrm{tr}(\mathbf{Q}_{s}) \leq P_{\textrm{tx}}, \mathbf{Q}_{s}\succeq 0, \textrm{rank}(\mathbf{Q}_{s})=1,\label{SEE_max_st_c5}  
		\end{align}
	\end{subequations}
	where $f_{s}(\mathbf{Q}_{s})$ and $g_{s}(\mathbf{Q}_{s})$ are defined as 
	\begin{align}
	f_{s}(\mathbf{Q}_{s})=\frac{\log_{2}(1\!+\!\frac{1}{\sigma_{su}^{2}}\mathbf{h}_{s}^{H}\mathbf{Q}_{s}\mathbf{h}_{s})\!-\!\log_{2}(1\!+\!\frac{1}{\sigma_{er}^{2}}\mathbf{h}_{e}^{H}\mathbf{Q}_{s}\mathbf{h}_{e})}{\textrm{tr}(\mathbf{Q}_{s})+P_{c}},\\
	g_{s}(\mathbf{Q}_{s})=\log_{2}(1\!+\!\frac{1}{\sigma_{su}^{2}}\mathbf{h}_{s}^{H}\mathbf{Q}_{s}\mathbf{h}_{s})\!-\!\log_{2}(1\!+\!\frac{1}{\sigma_{er}^{2}}\mathbf{h}_{e}^{H}\mathbf{Q}_{s}\mathbf{h}_{e}).
	\end{align}	
	In (\ref{SEE_max_st_c1})-(\ref{SEE_max_st_c3}), the parameters $\alpha$, $\beta$ and $\gamma$ are chosen to define the outage probabilities for SEE, secrecy rate and EH at ER as $0<\alpha<0.5$, $0<\beta<0.5$ and $0<\gamma<0.5$, respectively. Furthermore, the left hand side (LHS) of the constraint in (\ref{SEE_max_st_c1}) can be formulated as 
	\begin{align}
	&\textrm{Pr}\bigg\{\log\frac{1+\frac{1}{\sigma_{su}^{2}}\mathbf{h}_{s}^{H}\mathbf{Q}_{s}\mathbf{h}_{s}}{1+\frac{1}{\sigma_{er}^{2}}\mathbf{h}_{e}^{H}\mathbf{Q}_{s}\mathbf{h}_{e}}\geq \delta(\textrm{tr}(\mathbf{Q}_{s})+P_{c})\bigg\}\nonumber\\
	&=\textrm{Pr}\bigg\{\frac{1}{\sigma_{er}^{2}}\mathbf{h}_{e}^{H}\mathbf{Q}_{s}\mathbf{h}_{e}\!\leq\! (1\!+\!\frac{1}{\sigma_{su}^{2}}\mathbf{h}_{s}^{H}\mathbf{Q}_{s}\mathbf{h}_{s})2^{\!-\!\delta(\textrm{tr}(\mathbf{Q}_{s})\!+\!P_{c})}\!-\!1\bigg\}\nonumber\\
	&\overset{(a)}{=}1-\textrm{exp}\bigg(\frac{1-(1+\frac{1}{\sigma_{su}^{2}}\mathbf{h}_{s}^{H}\mathbf{Q}_{s}\mathbf{h}_{s})2^{-\delta(\textrm{tr}(\mathbf{Q}_{s})+P_{c})}}{\frac{1}{\sigma_{er}^{2}}\textrm{tr}(\mathbf{G}_{e}\mathbf{Q}_{s})}\bigg), \label{SEE_max_st_c1new}
	\end{align}
	where equality (a) in (24) is derived based on the fact that the random variable $\mathbf{h}_{e}^{H}\mathbf{Q}_{s}\mathbf{h}_{e}$ follows an exponential distribution with mean $\textrm{tr}(\mathbf{G}_{e}\mathbf{Q}_{s})$.		 
	Similarly, the LHS of the constraints in (\ref{SEE_max_st_c2}) and (\ref{SEE_max_st_c3}) can be recast respectively as
	\begin{align}
	&\textrm{(\ref{SEE_max_st_c2})}\Rightarrow 1-\textrm{exp}\bigg(\frac{1-(1+\frac{1}{\sigma_{su}^{2}}\mathbf{h}_{s}^{H}\mathbf{Q}_{s}\mathbf{h}_{s})2^{-R_{d}}}{\frac{1}{\sigma_{er}^{2}}\textrm{tr}(\mathbf{G}_{e}\mathbf{Q}_{s})}\bigg)\geq 1-\beta,\label{SEE_max_st_c2new}\\
	&\textrm{(\ref{SEE_max_st_c3})}\Rightarrow 1-\textrm{exp}\bigg(\frac{-\omega_{s}}{\zeta_{eh}\textrm{tr}(\mathbf{G}_{e}\mathbf{Q}_{s})}\bigg)\leq \gamma.\label{SEE_max_st_c3new}
	\end{align}
	Hence, the original problem in (\ref{SEE_max_st_ori}) can be rewritten as
	\begin{align}
	\max_{\mathbf{Q}_{s}} &~\delta \nonumber \\
	s.t.~&\textrm{exp}\bigg(\frac{1-(1+\frac{1}{\sigma_{su}^{2}}\mathbf{h}_{s}^{H}\mathbf{Q}_{s}\mathbf{h}_{s})2^{-\eta(\textrm{tr}(\mathbf{Q}_{s})+P_{c})}}{\frac{1}{\sigma_{er}^{2}}\textrm{tr}(\mathbf{G}_{e}\mathbf{Q}_{s})}\bigg)\leq \alpha,\nonumber\\
	& \textrm{exp}\bigg(\frac{1-(1+\frac{1}{\sigma_{su}^{2}}\mathbf{h}_{s}^{H}\mathbf{Q}_{s}\mathbf{h}_{s})2^{-R_{d}}}{\frac{1}{\sigma_{er}^{2}}\textrm{tr}(\mathbf{G}_{e}\mathbf{Q}_{s})}\bigg)\leq \beta,\nonumber\\
	&1-\textrm{exp}\bigg(\frac{-\omega_{s}}{\zeta_{eh}\textrm{tr}(\mathbf{G}_{e}\mathbf{Q}_{s})}\bigg)\leq\gamma,~\textrm{(\ref{SEE_max_st_c4}),~(\ref{SEE_max_st_c5})}.  
	\end{align}
	By performing some matrix manipulations, the following equivalent optimization problem is derived:
	\begin{subequations}\label{SEE_max_st_eq}
		\begin{align}
		\max_{\mathbf{Q}_{s}} &\frac{\log_{2}(1\!+\!\frac{1}{\sigma_{su}^{2}}\mathbf{h}_{s}^{H}\mathbf{Q}_{s}\mathbf{h}_{s})\!-\!\log_{2}(1\!-\!\frac{1}{\sigma_{su}^{2}}\textrm{tr}(\mathbf{G}_{e}\mathbf{Q}_{s})\ln\alpha)}{\textrm{tr}(\mathbf{Q}_{s})+P_{c}}  \\
		s.t.~&1+\frac{1}{\sigma_{su}^{2}}\mathbf{h}_{s}^{H}\mathbf{Q}_{s}\mathbf{h}_{s}\geq 2^{R_{d}} \label{SEE_max_st_eqc1} -\frac{2^{R_{d}}}{\sigma_{er}^{2}}\textrm{tr}(\mathbf{G}_{e}\mathbf{Q}_{s})\ln\beta,\\
		&\zeta_{eh}\textrm{tr}(\mathbf{G}_{e}\mathbf{Q}_{s})\ln(1-\gamma)\leq -\omega_{s}, \label{SEE_max_st_eqc2}\\
		&\textrm{(\ref{SEE_max_st_c4}), (\ref{SEE_max_st_c5})}. 
		\end{align}
	\end{subequations}
	In the following subsections, we present an approach to solve this equivalent problem based on non-linear fractional and DC programming. 
	
	\subsection{Non-linear Fractional Programming}
	Following a similar approach as in (13), we first define a parametric problem w.r.t. $\lambda$, as follows:
	\begin{align} \label{SEE_max_st_nl}
	F_{s}(\lambda)=&\max_{\mathbf{Q}_{s}}~\log_{2}(1+\frac{1}{\sigma_{su}^{2}}\mathbf{h}_{s}^{H}\mathbf{Q}_{s}\mathbf{h}_{s})\nonumber\\&~-\log_{2}(1-\frac{1}{\sigma_{er}^{2}}\textrm{tr}(\mathbf{G}_{e}\mathbf{Q}_{s})\ln\alpha)-\lambda( \textrm{tr}(\mathbf{Q}_{s})+P_{c}),\nonumber\\
	s.t.~& \textrm{(28b), (28c), (21e), (21f)}. 
	\end{align}
	The value of $\lambda_{i}$ at the $i$th iteration can be chosen as the SEE obtained at the previous iteration. Furthermore, $\lambda_{i}$ should be updated at each iteration as $
	\lambda_{i+1}=\frac{R_{st}^{i}(\mathbf{Q}_{s})}{P_{st}^{i}(\mathbf{Q}_{s})}
	$, 	where $R_{st}^{i}(\mathbf{Q}_{s})$ and $P_{st}^{i}(\mathbf{Q}_{s})$ denote the achieved secrecy rate and total power consumption that are obtained by solving  (\ref{SEE_max_st_nl}) for the given $\lambda_{i}$, respectively. In practice, the iterative process will be repeated until the following inequality is satisfied:
	
	\begin{align}
	\Delta F_{s}(\lambda)=|R_{s}^{i}(\mathbf{Q}_{s})-\lambda_{i}P_{t}^{i}(\mathbf{Q}_{s})|\leq \varsigma,\label{st_statisfy} 
	\end{align}
	with a small convergence tolerance $\varsigma>0$. 
	
	\subsection{DC Programming}
	Next, we exploit DC programming to solve the optimization problem defined in (29). Based on this approximation, the problem can be converted into the following equivalent problem:
	\begin{align}\label{SEE_relax}
	\max_{\mathbf{Q}_{s}}&\{\log_{2}(\!1\!+\!\frac{1}{\sigma_{su}^{2}}\mathbf{h}_{s}^{H}\mathbf{Q}_{s}\mathbf{h}_{s}\!)\!-\!f_{s}(\mathbf{Q}_{s},\mathbf{Q}_{s}^{k})\!-\!\lambda_{i}[ \textrm{tr}(\mathbf{Q}_{s})\!+\!P_{c}]\}\nonumber\\
	s.t.~& \textrm{(28b), (28c), (21e), (21f)}, 
	\end{align}
	\noindent where 
	\begin{align}
	f_{s}(\mathbf{Q}_{s},\mathbf{Q}_{s}^{k})=\log_{2}(1-\frac{1}{\sigma_{er}^{2}}\textrm{tr}(\mathbf{G}_{e}\mathbf{Q}_{s}^{k})\ln\alpha))-\nonumber\\
	\frac{\frac{\ln\alpha}{\sigma_{er}^{2}}\textrm{tr}(\mathbf{G}_{e}(\mathbf{Q}_{s}-\mathbf{Q}_{s}^{k}))}{(1-\frac{1}{\sigma_{er}^{2}}\textrm{tr}(\mathbf{G}_{e}\mathbf{Q}_{s}^{k})\ln\alpha)\ln2}. 
	\end{align}
	Then, by exploiting SDR, the relaxed problem can be defined by dropping the rank-one constraint $\textrm{rank}(\mathbf{Q}_{s}) = 1$ in (31) as follows:
	\begin{align}\label{SEE_max_st_dc}
	\max_{\mathbf{Q}_{s}}&\{\log_{2}(\!1\!+\!\frac{1}{\sigma_{su}^{2}}\mathbf{h}_{s}^{H}\mathbf{Q}_{s}\mathbf{h}_{s})\!-\!f_{s}(\mathbf{Q}_{s},\mathbf{Q}_{s}^{k})\!-\!\lambda_{i}[ \textrm{tr}(\mathbf{Q}_{s})\!+\!P_{c}]\}\nonumber\\
	s.t.~& \textrm{(28b), (28c), (21e)},\textrm{tr}(\mathbf{Q}_{s}) \leq P_{\textrm{tx}}, \mathbf{Q}_{s}\succeq 0. 
	\end{align}

	\begin{Proposition}\label{proposition:rank_proof2}
		Provided that the problem (\ref{SEE_max_st_dc}) is feasible, the optimal solution will be always rank-one.
	\end{Proposition}
	\begin{IEEEproof}
		Please refer to Appendix \ref{proof_of_proposition2}.
	\end{IEEEproof}
	~\\
	
	This approximated problem is convex in terms of $\mathbf{Q}_{s}$. Hence, the suboptimal solution $\mathbf{Q}_{s}^{*}$ of the original problem in (\ref{SEE_max_st_ori}) can be obtained through iteratively solving problem (\ref{SEE_max_st_dc}) and updating $\mathbf{Q}_{s}^{k}$ at each iteration based on the solution obtained from the previous iteration. The proposed algorithm based on DC programming is provided in Algorithm 2.
	\begin{algorithm}
		\caption{\!\!: Iterative algorithm for solving (\ref{SEE_max_st_dc})}\label{alg:stDC}
		\begin{algorithmic}[1]
			\State Initialize $i=0$ and choose an initial value $\lambda_{0}$ for $\lambda$;
			\Repeat $\leftarrow$ Outer loop (Dinkelbach's algorithm)
			\State Initialize $k=0$, choose an initial value $\mathbf{Q}_{s}^{k}=0$ and \phantom a\phantom a\phantom . $F_{s}(\lambda)^{i,k}=0$;
			\Repeat $\leftarrow$ Inner loop (DC programming)
			\State Solve the problem (\ref{eq:SEE_max_relaxed}) with $\lambda=\lambda_{i}$ and obtain \phantom a\phantom a\phantom a\phantom a\phantom a\phantom a\phantom .$\mathbf{Q}_{s}^{k+1}$;
			\State Compute  $\!F_{s}(\lambda)^{i,k+1}\!=\!\log_{2}(1\!+\!\frac{1}{\sigma_{su}^{2}}\mathbf{h}_{s}^{H}\mathbf{Q}_{s}^{k+1}$$\mathbf{h}_{s})-\phantom a\phantom a\phantom a\phantom a\phantom a\phantom .\log_{2}(1-\frac{1}{\sigma_{er}^{2}}\textrm{tr}(\mathbf{G}_{e}\mathbf{Q}_{s}^{k+1})\ln\alpha)- \lambda_{i}( \textrm{tr}(\mathbf{Q}_{s}^{k+1})+\phantom a\phantom a\phantom a\phantom a\phantom a\phantom a P_{c})$;
			\State $\Delta\mu=F_{s}(\lambda)^{i,k+1}-F(\lambda)^{i,k}$;
			\State Update $k=k+1$;
			\Until $|\Delta \mu|\leq \zeta$;
			\State Update $\lambda_{i+1}=\frac{R_{st}^{i}(\mathbf{Q}_{s})}{P_{st}^{i}(\mathbf{Q}_{s})}$;		
			\State $i=i+1$;
			\Until (\ref{st_statisfy}) satisfied;
			\State Return $\lambda^{*}=\lambda_{i}, P^{*}_{st}(\mathbf{Q}_{s})=P_{st}^{i-1}(\mathbf{Q}_{s}), R_{st}^{*}(\mathbf{Q}_{s})=R_{st}^{i-1}(\mathbf{Q}_{s})$;
			\State Output $\lambda_{i}$.
		\end{algorithmic}
	\end{algorithm}
	\subsection{Computational Complexity}
The problem defined in (32) has four linear constraints, and one LMI constraint of size $N_{T}$. By utilizing the same method as in the literature \cite{wang2014outage,ben2001lectures}, the computational complexity of the solution for~(21) should be in the order of $TK\sqrt{N_{T}+4}[n^{3}+n(N_{T}^{3}+4)+n^{2}(N_{T}^{2}+4)]\ln(1/\epsilon),$ where $T$ and $K$ are the numbers of iterations of the inner and outer loop, respectively.

	\section{SEE Maximization with Imperfect CSI}
	\label{sec:see_max_icsi}

	In this section, we develop a tractable approach to solve the robust SEE maximization problem with imperfect CSI on all set of channels. First we reformulate this robust problem into a simple parametric problem by exploiting a well-known non-linear fractional programming. Then, we show that the optimal solution can be obtained by solving a series of SDP. The actual channel coefficients can be modelled by channel uncertainties as follows:
	\begin{align}
	\mathbf{h}_{s}=\mathbf{\hat{h}}_{s}+\mathbf{\hat{e}}_{s},~\mathbf{h}_{p}=\mathbf{\hat{h}}_{p}+\mathbf{\hat{e}}_{p},~\mathbf{h}_{e}=\mathbf{\hat{h}}_{e}+\mathbf{\hat{e}}_{e},
	\end{align}
	where $\mathbf{\hat{e}}_{s}$, $\mathbf{\hat{e}}_{p}$ and $\mathbf{\hat{e}}_{e}$ represent the channel uncertainties. Furthermore, these channel uncertainties can be defined based on an ellipsoid model as
	\begin{align}
	||\mathbf{\hat{e}}_{s}||=||\mathbf{h}_{s}-\mathbf{\hat{h}}_{s}||\leq\epsilon_{s},\\
	||\mathbf{\hat{e}}_{p}||=||\mathbf{h}_{p}-\mathbf{\hat{h}}_{p}||\leq\epsilon_{p},\\
	||\mathbf{\hat{e}}_{e}||=||\mathbf{h}_{e}-\mathbf{\hat{h}}_{e}||\leq\epsilon_{e},
	\end{align}
	where $\epsilon_{s}\geq0$, $\epsilon_{p}\geq0$ and $\epsilon_{e}\geq0$ are Euclidean norm-based error bounds. By incorporating these channel uncertainties in the original design, the SEE maximization problem can be reformulated into the following robust optimization framework:
	\begin{subequations}\label{c}
		\begin{align}
		\max_{\mathbf{Q}_{s}}&~ \frac{R_{m}(\mathbf{Q}_{s})}{P_{m}(\mathbf{Q}_{s})} \label{eq:SEE_max_imreobj} \\
		s.t.&~ \log_{2}(1+\frac{1}{\sigma_{su}^{2}}(\mathbf{\hat{h}}_{s}+\mathbf{\hat{e}}_{s})^{H}\mathbf{Q}_{s}(\mathbf{\hat{h}}_{s}+\mathbf{\hat{e}}_{s}))-\log_{2}(1+\nonumber\\
		&~\frac{1}{\sigma_{er}^{2}}(\mathbf{\hat{h}}_{e}+\mathbf{\hat{e}}_{e})^{H}\mathbf{Q}_{s}(\mathbf{\hat{h}}_{e}+\mathbf{\hat{e}}_{e}))\geq R_{d}, \label{eq:SEE_Max_imrec1}\\
		&~\zeta_{eh}(\mathbf{\hat{h}}_{e}+\mathbf{\hat{e}}_{e})^{H}\mathbf{Q}_{s}(\mathbf{\hat{h}}_{e}+\mathbf{\hat{e}}_{e}) \geq \omega_{s}, \label{eq:SEE_Max_imrec2}\\
		&~ (\mathbf{\hat{h}}_{p}+\mathbf{\hat{e}}_{p})^{H}\mathbf{Q}_{s}(\mathbf{\hat{h}}_{p}+\mathbf{\hat{e}}_{p})\leq P_{f}, \label{eq:SEE_Max_imrec4}\\
		&~ \textrm{tr}(\mathbf{Q}_{s}) \leq P_{\textrm{tx}}, \mathbf{Q}_{s}\succeq \mathbf{0}, \textrm{rank}(\mathbf{Q}_{s})=1,  \label{eq:SEE_Max_imrec3}
		\end{align}
	\end{subequations}
	where $R_{m}(\mathbf{Q}_{s})=\log_{2}(1+\frac{1}{\sigma_{su}^{2}}(\mathbf{\hat{h}}_{s}+\mathbf{\hat{e}}_{s})^{H}\mathbf{Q}_{s}(\mathbf{\hat{h}}_{s}+\mathbf{\hat{e}}_{s}))-\log_{2}(1+\frac{1}{\sigma_{er}^{2}}(\mathbf{\hat{h}}_{e}+\mathbf{\hat{e}}_{e})^{H}\mathbf{Q}_{s}(\mathbf{\hat{h}}_{e}+\mathbf{\hat{e}}_{e}))$ and $P_{m}=\textrm{tr}(\mathbf{Q}_{s})+P_{c}$. Next we present a two stage reformulation to solve the above robust problem.
	
	\subsection{A Two-stage Reformulation of Problem (\ref{c})}
	First, we exploit Theorem 1, Lemma 1 and Lemma 2 to recast the problem in (\ref{c}) into the following parametric problem:
	\begin{align}\label{SEE_max_im}
	&\max_{\mathbf{Q}_{s}}~[R_{m}(\mathbf{Q}_{s})-\lambda P_{m}(\mathbf{Q}_{s})]\nonumber\\
	&~s.t. ~\textrm{(\ref{eq:SEE_Max_imrec1})-(\ref{eq:SEE_Max_imrec3})}. 
	\end{align}
	It is obvious that the original problem (\ref{c}) is transformed into a parameterized polynomial subtractive form. As a result, the original problem is reformulated to determine $\lambda^{*}$ and $\mathbf{Q}_{s}^{*}$ to satisfy the condition defined in (\ref{conditional1}). Furthermore, we exploit Dinklebach's algorithm, which iteratively solves the problem defined in (\ref{SEE_max_im}) for a given $\lambda$ and updates $\lambda$ in the next iteration as follows: 
	\begin{align}
	\lambda_{i+1}=\frac{R_{m}^{i}(\mathbf{Q}_{s})}{P_{m}^{i}(\mathbf{Q}_{s})},
	\end{align}
	where $R_{m}^{i}(\mathbf{Q}_{s})$ and $P_{m}^{i}(\mathbf{Q}_{s})$ denote the achieved secrecy rate and total power consumption of (\ref{c}) for the given $\lambda_{i}$, respectively, and subscript $i$ denotes the iteration number. This iterative process will be terminated once the following condition is satisfied:
	\begin{align}
	\Delta F_{m}(\lambda)=|R_{m}^{i}(\mathbf{Q}_{s})-\lambda_{i}P_{m}^{i}(\mathbf{Q}_{s})|\leq\rho , \label{im_satisfy}
	\end{align}
	where $\rho>0$ is the termination threshold. By introducing a new slack variable $\tau$, the following equivalent problem of (\ref{SEE_max_im}) is derived:
	\begin{subequations}\label{SEE_max_imperfect}
		\begin{align}
		F_{m}(\lambda)=\max_{\mathbf{Q}_{s},\tau}&~\log_{2}\bigg(\frac{1+\frac{1}{\sigma_{su}^{2}}(\mathbf{\hat{h}}_{s}+\mathbf{\hat{e}}_{s})^{H}\mathbf{Q}_{s}(\mathbf{\hat{h}}_{s}+\mathbf{\hat{e}}_{s})}{\tau}\bigg)-\nonumber\\
		&~\lambda[ \textrm{tr}(\mathbf{Q}_{s})+P_{c}]\}\nonumber\\
		s.t. ~&1+\frac{1}{\sigma_{er}^{2}}(\mathbf{\hat{h}}_{e}+\mathbf{\hat{e}}_{e})^{H}\mathbf{Q}_{s}(\mathbf{\hat{h}}_{e}+\mathbf{\hat{e}}_{e})\leq\tau,\label{SEE_max_imperfect_c1}\\
		~&\log_{2}(1+\frac{1}{\sigma_{su}^{2}}(\mathbf{\hat{h}}_{s}+\mathbf{\hat{e}}_{s})^{H}\mathbf{Q}_{s}(\mathbf{\hat{h}}_{s}+\mathbf{\hat{e}}_{s}))-\nonumber\\
		~&\log_{2}(\tau)\!\geq\! R_{d},\label{SEE_max_imperfect_c2}\\
		~&\textrm{(\ref{eq:SEE_Max_imrec2})-(\ref{eq:SEE_Max_imrec3})}. \label{SEE_max_imperfect_c3}
		\end{align}
	\end{subequations}
	Next, by introducing a new slack variable $\nu$, the above problem is transformed into a two-stage problem, namely, outer problem and inner problem. The outer problem can be defined as
	\begin{align}\label{SEE_max_imperfect_outer}
	F_{m}(\lambda)=&\max_{\nu} ~\{\log_{2}\Theta(\lambda,\nu)-\nu\}\nonumber\\
	&s.t.~\lambda(\nu_{min}+P_{c})\leq\nu\leq\lambda(P_{\textrm{tx}}+P_{c}),
	\end{align}
	where $\nu_{min}$ denotes the minimum value of $\nu$ which can be obtained by solving the power minimization problem defined in (\ref{powermin}). Furthermore, the inner problem determines $\Theta(\lambda,\nu)$ for a given $\nu$, which can be defined as
	\begin{align}\label{SEE_max_imperfect_inner}
	\Theta(\lambda,\nu)=\max_{\mathbf{Q}_{s},\tau}&~\frac{1+\frac{1}{\sigma_{su}^{2}}(\mathbf{\hat{h}}_{s}+\mathbf{\hat{e}}_{s})^{H}\mathbf{Q}_{s}(\mathbf{\hat{h}}_{s}+\mathbf{\hat{e}}_{s})}{\tau}\nonumber\\
	s.t. &~\lambda(\textrm{tr}(\mathbf{Q}_{s})+P_{c})\leq \nu,~\textrm{(\ref{SEE_max_imperfect_c1})-(\ref{SEE_max_imperfect_c3})}. 
	\end{align}
	However, the inner problem (\ref{SEE_max_imperfect_inner}) is a single variable optimization problem with a set of feasible values $\nu\in [\lambda(\nu_{min}+P_{c}),\lambda(P_{\textrm{tx}}+P_{c})]$. It is well-known that the optimal value can be efficiently obtained through an one-dimensional search. By relaxing the non-convex rank-one constraint, the relaxed inner problem can be defined as
	\begin{align}\label{SEE_max_imperfecr_inner}
	\Theta(\lambda,\nu)=\max_{\mathbf{Q}_{s},\tau}&~\frac{1+\frac{1}{\sigma_{su}^{2}}(\mathbf{\hat{h}}_{s}+\mathbf{\hat{e}}_{s})^{H}\mathbf{Q}_{s}(\mathbf{\hat{h}}_{s}+\mathbf{\hat{e}}_{s})}{\tau}\nonumber\\
	s.t.&~\textrm{(\ref{SEE_max_imperfect_c1}),(\ref{SEE_max_imperfect_c2}),(\ref{eq:SEE_Max_imrec2}),(\ref{eq:SEE_Max_imrec4})},\nonumber\\
	&~\lambda(\textrm{tr}(\mathbf{Q}_{s})+P_{c})\leq \nu, \textrm{tr}(\mathbf{Q}_{s}) \leq P_{\textrm{tx}}, \mathbf{Q}_{s}\succeq \mathbf{0}.
	\end{align}
	\begin{Proposition}\label{proposition:rank_proof3}
		Provided that the problem (\ref{SEE_max_imperfecr_inner}) is feasible, the optimal solution will be always rank-one.
	\end{Proposition}
	\begin{IEEEproof}
		Please refer to Appendix \ref{proof_of_proposition3}.
	\end{IEEEproof}
	~\\
	
	In the following subsection, we present an SDP formulation based on the Charnes-Cooper transformation \cite{charnes1962programming} and S-Procedure \cite{boyd1994linear} to solve the above relaxed inner problem.
	
	\subsection{SDP-based Reformulation of the Inner Problem (\ref{SEE_max_imperfecr_inner})}
	
	First, we introduce the Charnes-Cooper transformation  \cite{charnes1962programming} as
	\begin{align}\label{CC_trans}
	\mathbf{W}=\frac{\mathbf{Q}_{s}}{\tau},\Gamma=\frac{1}{\tau},
	\end{align}
	and a slack variable $\varrho$ to recast the inner problem (\ref{SEE_max_imperfecr_inner}) as
	\begin{subequations}\label{SEE_max_csierr}
		\begin{align}
		\Theta(\lambda,\nu)&=\max_{\mathbf{W},\Gamma}~\varrho\nonumber\\
		s.t.&~\Gamma+\frac{1}{\sigma_{su}^{2}}(\mathbf{\hat{h}}_{s}+\mathbf{\hat{e}_{s}})^{H}\mathbf{W}(\mathbf{\hat{h}}_{s}+\mathbf{\hat{e}_{s}})\geq \varrho,\\
		&~\Gamma+\frac{1}{\sigma_{er}^{2}}(\mathbf{\hat{h}}_{e}+\mathbf{\hat{e}_{e}})^{H}\mathbf{W}(\mathbf{\hat{h}}_{s}+\mathbf{\hat{e}_{s}})\leq 1,\\
		&~\Gamma+\frac{1}{\sigma_{su}^{2}}(\mathbf{\hat{h}}_{s}+\mathbf{\hat{e}_{s}})^{H}\mathbf{W}(\mathbf{\hat{h}}_{s}+\mathbf{\hat{e}_{s}})\geq 2^{R_{d}},\\
		&~\zeta_{eh}(\mathbf{\hat{h}}_{e}+\mathbf{\hat{e}_{e}})^{H}\mathbf{W}(\mathbf{\hat{h}}_{e}+\mathbf{\hat{e}_{e}}) \geq \omega_{s}\Gamma,\\
		&~ (\mathbf{\hat{h}}_{p}+\mathbf{\hat{e}_{p}})^{H}\mathbf{W}(\mathbf{\hat{h}}_{p}+\mathbf{\hat{e}_{p}})\leq \Gamma P_{f}, \\
		&~\lambda(\textrm{tr}(\mathbf{W})+\Gamma P_{c})\leq\Gamma \nu,\textrm{tr}(\mathbf{W}) \leq \Gamma P_{\textrm{tx}}, \mathbf{W}\succeq \mathbf{0}. \label{SEE_max_im_cx}
		\end{align}
	\end{subequations}
	To further proceed with this problem, the following lemma is required:
	\begin{Lemma} (\emph{S-Procedure}~\cite{boyd1994linear}) Define $f_{i}(\mathbf{x})$, $i=1,2$ such as
		\begin{align} f_{i}(\mathbf{x})=\mathbf{x}^{H}\mathbf{A}_{i}\mathbf{x}+2Re\{\mathbf{b}_{i}^{H}\mathbf{x}\}+c_{i},
		\end{align}
		in which $\mathbf{x}\in\mathcal{C}^{N_{T}\times1}$, $\mathbf{A}_{i}\in\mathcal{C}^{N_{T}\times N_{T}}$, $\mathbf{b}_{i}\in\mathcal{C}^{N_{T}\times1}$ and $c_{i}\in\mathcal{R}$. The implication $f_{1}(\mathbf{x})\geq 0 \rightarrow f_{2}(\mathbf{x})\geq 0$ holds if and only if there exists a $\vartheta\geq 0$ such that
		\begin{align}
		\left[
		\begin{matrix}
		\mathbf{A}_{2} & \mathbf{b}_{2}\\
		\mathbf{b}_{2}^{H} & c_{2}\\
		\end{matrix}
		\right]
		-
		\vartheta\left[
		\begin{matrix}
		\mathbf{A}_{1} & \mathbf{b}_{1}\\
		\mathbf{b}_{1}^{H} & c_{1}\\
		\end{matrix}
		\right]
		\succeq \mathbf{0}. 
		\end{align}
	\end{Lemma}
	By applying Lemma 3, the inner problem defined in (\ref{SEE_max_csierr}) can be recast as
	\begin{subequations}\label{SEE_max_im_sp}
		\begin{align}
		\Theta(\lambda,\nu)&=\max_{\mathbf{W},\tau}~\varrho\\
		s.t.&~\left[
		\begin{matrix}
		\vartheta_{1}\mathbf{I}+\frac{1}{\sigma_{su}^{2}}\mathbf{W} & \frac{1}{\sigma_{su}^{2}}\mathbf{W}\mathbf{\hat{h}}_{s}\\
		\frac{1}{\sigma_{su}^{2}}\mathbf{\hat{h}}_{s}^{H}\mathbf{W} & t_{1}
		\end{matrix}
		\right]\succeq \mathbf{0},\\
		&~\left[
		\begin{matrix}
		\vartheta_{2}\mathbf{I}-\frac{1}{\sigma_{er}^{2}}\mathbf{W} & -\frac{1}{\sigma_{er}^{2}}\mathbf{W}\mathbf{\hat{h}}_{e}\\
		-\frac{1}{\sigma_{er}^{2}}\mathbf{\hat{h}}_{e}^{H}\mathbf{W} & t_{2}
		\end{matrix}
		\right]\succeq \mathbf{0},\label{constraint_1}\\
		&~\left[
		\begin{matrix}
		\vartheta_{3}\mathbf{I}+\frac{1}{\sigma_{su}^{2}}\mathbf{W} & \frac{1}{\sigma_{su}^{2}}\mathbf{W}\mathbf{\hat{h}}_{s}\\
		\frac{1}{\sigma_{su}^{2}}\mathbf{\hat{h}}_{s}^{H}\mathbf{W} & 
		t_{3}
		\end{matrix}
		\right]\succeq \mathbf{0},\\
		&~\left[
		\begin{matrix}
		\vartheta_{4}\mathbf{I}+\zeta_{eh}\mathbf{W} & \zeta_{eh}\mathbf{W}\mathbf{\hat{h}}_{e}\\
		\zeta_{eh}\mathbf{\hat{h}}_{e}^{H}\mathbf{W} & 
		t_{4}
		\end{matrix}
		\right]\succeq \mathbf{0},\\
		&~\left[
		\begin{matrix}
		\vartheta_{5}\mathbf{I}-\mathbf{W} & -\mathbf{W}\mathbf{\hat{h}}_{p}\\
		-\mathbf{\hat{h}}_{p}^{H}\mathbf{W} & t_{5}
		\end{matrix}
		\right]\succeq \mathbf{0},\label{constraint_4}\\
		&~~\textrm{(\ref{SEE_max_im_cx})}, \vartheta_{1}\geq 0, \vartheta_{2}\geq 0, \vartheta_{3}\geq 0, \vartheta_{4}\geq 0, \vartheta_{5}\geq 0, 
		\end{align}
	\end{subequations}
	where $t_{1}=\frac{1}{\sigma_{su}^{2}}\mathbf{\hat{h}}_{s}^{H}\mathbf{W}\mathbf{\hat{h}}_{s}+\Gamma-\varrho-\vartheta_{1}\epsilon_{s}^{2}$, $t_{2}=1-\frac{1}{\sigma_{er}^{2}}\mathbf{\hat{h}}_{e}^{H}\mathbf{W}\mathbf{\hat{h}}_{e}-\Gamma-\vartheta_{2}\epsilon_{e}^{2}$, $t_{3}=\frac{1}{\sigma_{su}^{2}}\mathbf{\hat{h}}_{s}^{H}\mathbf{W}\mathbf{\hat{h}}_{s}-2^{R_{d}}+\Gamma-\vartheta_{3}\epsilon_{s}^{2}$, $t_{4}=\zeta_{eh}\mathbf{\hat{h}}_{e}^{H}\mathbf{W}\mathbf{\hat{h}}_{e}-\omega_{s}\Gamma-\vartheta_{4}\epsilon_{e}^{2}$ and $t_{5}=\Gamma P_{f}-\mathbf{\hat{h}}_{p}^{H}\mathbf{W}\mathbf{\hat{h}}_{p}-\vartheta_{5}\epsilon_{p}^{2}$. This inner problem in (\ref{SEE_max_im_sp}) is convex and can be efficiently solved through existing convex optimization software~\cite{boyd2004convex}. Furthermore, $\upsilon_{min}$ can be obtained through solving the following convex problem
	\begin{align}\label{powermin}
	\upsilon_{min}=\min_{\mathbf{W},\Gamma}&~\textrm{tr}(\mathbf{W})\nonumber\\
	s.t.&~(\textrm{\ref{constraint_1}})-(\textrm{\ref{constraint_4}}),~\textrm{(\ref{SEE_max_im_cx})}, \nonumber\\
	&~\vartheta_{2}\geq 0, \vartheta_{3}\geq 0, \vartheta_{4}\geq 0, \vartheta_{5}\geq 0. 
	\end{align}
	The above developed procedures for solving this robust SEE maximization problem are summarized in Algorithm 3.
	\begin{algorithm}
		\caption{\!\!: Dinkelbach's algorithm for solving the SEE maximization problem in (\ref{c}) }\label{alg:imD}
		\begin{algorithmic}[1]
			\State Initialization $i=0$ and $\lambda =\lambda_{0}$;
			\Repeat
			\State Perform a one-dimensional search over $\nu$ to obtain  \phantom a \phantom a   \phantom a the optimal values $(\nu^{*},\Theta(\lambda_{i},\nu^{*}))$ for the outer  \phantom a \phantom a  \phantom a \phantom a problem  in (\ref{SEE_max_imperfect_outer}), where each $\Theta(\lambda_{i},\nu)$ is obtained by  \phantom a \phantom a  solving the inner problem in (\ref{SEE_max_im_sp});
			\State Retrieve the corresponding $\mathbf{Q}_{s}^{*}$ through (\ref{CC_trans});
			\State Compute $F(\lambda_{i})=R_{m}^{i}(\mathbf{Q}_{s}^{*})-\lambda_{i}P^{i}_{m}(\mathbf{Q}_{s}^{*})$;
			\State Update $\lambda_{i+1}$ by using (43);
			\State $i=i+1$;
			\Until the condition in (\ref{im_satisfy}) is satisfied;
			\State Return the optimal value of the inner problem in (\ref{SEE_max_im_sp}) to the outer problem in (\ref{SEE_max_imperfect_outer});
			\State Output the maximum SEE obtained by the outer problem in (\ref{SEE_max_imperfect_outer}).
		\end{algorithmic}
	\end{algorithm}
\subsection{Computational Complexity}
The problem defined in (49) has five LMI constraints of size $N_{T}+1$, one LMI constraint of size $N_{T}$, and seven linear constraints. The number of variables $n$ is in the order of $N_{T}^{2}$. By following the same method as in \cite{wang2014outage, ben2001lectures}, the computational complexity of solving the problem defined in (37) should be on the order of $TK\sqrt{6N_{T}+12}\{n^{3}+n[5(N_{T}+1)^{3}+N_{T}^{3}+7]+n^{2}[5(N_{T}+1)^{2}+N_{T}^{2}+7]\}\ln(1/\epsilon),$ where $T$ and $K$ are the numbers of iterations of the one-dimensional search and Dinkelbach's algorithm, respectively.
	\section{Simulation Results}
	\label{sec:results}
	
	To validate the proposed approaches, we perform a number of
        simulations based on the model we adopted, but with different
        configurations. We consider an underlay MISO CR network where
        the multi-antenna SU-Tx simultaneously transmits confidential
        information and energy to the SU-Rx and ER, respectively. This
        communication is established by sharing the spectrum that has
        been assigned for signal transmission between a PU-Tx and a
        PU-Rx. It is assumed that the SU-Tx is equipped with three
        ($N_{t} = 3$) antennas, while the PU-Rx, SU-Rx and ER each has
        a single antenna.  All channel coefficients are
          modelled by taking both large- and small-scale fading
          effects into account. The path loss coefficients caused by
          large-scale fading are modelled as $\sqrt{d_{i}^{-\alpha}}$,
          where $d_{i}$ is the distance between the SU-Tx and user
          $i$, and $\alpha=2.7$ denotes the path loss exponent. We
          employ Rayleigh fading as the small-scale fading, which can
          be modelled as
          $\chi_{i}\sim\mathcal{CN}(0,\mathbf{I})$. Therefore, the
          channel coefficients are formulated as $\mathbf{h}_{i}=
          \mathbf{\chi_{i}}\sqrt{d_{i}^{-\alpha}},~i=s,e,p$. We assume
          that all distances between the SU-TX and three terminals
          are the same and normalized to unit
          value.\footnote{Similar assumptions have been
            employed in the literature
            \cite{zhong2014wireless,pei2010secure}. In practical
            communication scenarios, this assumption may be
            implemented in wireless sensors and mobile wearable
            devices \cite{visser2013rf, sun2013wireless}.} The maximum
        interference leakage to the PU-Rx is set to be 0
        dBW. Furthermore, the energy conversion ratio is assumed to be
        $\zeta_{eh}= 0.5$. The noise variances at the SU-Rx and ER are
        set to $\sigma_{su}^{2}=1$ and $\sigma_{er}^{2}=1$,
        respectively. The convergence tolerances $\varepsilon$,
        $\varsigma$ and $\rho$ defined in (\ref{convergence_V1}),
        (\ref{st_statisfy}) and (\ref{im_satisfy}), respectively, are
        assumed to be $10^{-3}$.
	
          Note that in majority of the simulation
          scenarios outlined here, the difference between the transmit
          power and the EH requirement is large, and thus, may raise
          concerns on the validity of this assumption. However, we
          believe this assumption being is reasonable, for the
          following two reasons: 1) similar assumptions (large gap
          between transmit power and EH requirements) have been widely
          made in the literature, for instance in~\cite{ni2018energy,
            zhu2018robust, zhu2017beamforming}; and 2) for the
          secondary system in CR network, the ER could be wireless
          sensors or Internet-of-Things devices. For these devices,
          only a power in the order of milliwatts is sufficient
          \cite{ng2019wireless}.
	
	First, we evaluate the convergence of the proposed algorithms by randomly generating different five set of channels for each CSI assumption (perfect CSI, statistical CSI on the ER channel and imperfect CSI). Figure 2 presents the convergence of the achieved SEE with perfect CSI obtained by Algorithm 1. Here,  the target secrecy rate $(R_{d})$, the transmit power consumption $(P_{\textrm{tx}})$ and the EH requirement $(\omega_{s})$ have been set to 0.5 bps/Hz, 20 dBW and $\omega_{s}=-20$ dBW, respectively. We then show the convergence of achieved SEE with statistical CSI on the ER's channel, and the convergence of the achieved SEE with imperfect CSI in Figure  3 and~4. Here, the system parameters are:  target secrecy rate $R_{d} = 0.5$ bps/Hz, the transmit power consumption $P_{\textrm{tx}}=20$ dBW and the EH requirement $\omega_{s}=-3$ dBW. As can be observed, the maximum SEE and the convergence of our proposed algorithms can be attained with a limited number of iterations.
	
\begin{center}
		\begin{figure}[t!]
			\centering\includegraphics[width=\linewidth]{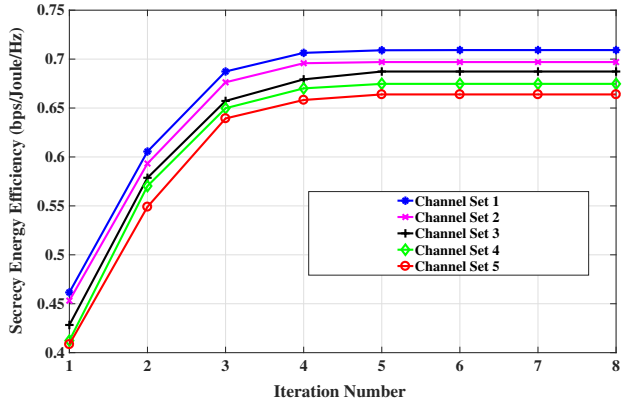}
			\caption{Convergence of the proposed algorithm with perfect CSI assumption in Algorithm 1 for different sets of channels.}
		\end{figure}
	\end{center}
	\begin{center}
		\begin{figure}[t!]
			\centering\includegraphics[width=\linewidth]{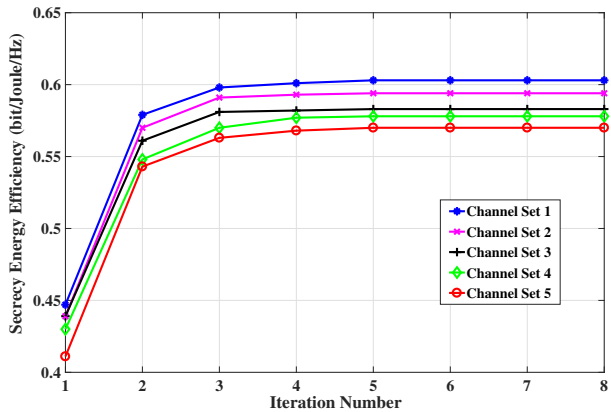}
			\caption{Convergence of the proposed algorithm with the assumption of statistical CSI on the ER channel in Algorithm 2 for different sets of channels.}
		\end{figure}
	\end{center}
\vspace{-4 em}
		     		     
	Figure 5 shows the achieved SEE with different target secrecy rates, and EH requirements under the assumption of perfect CSI. As seen in Figure  5, the optimal SEE decreases as the target rate increases. Note that zero SEE means that the original SEE maximization problem is infeasible; this is due to unachievable target secrecy rate constraint. Furthermore, it is possible to achieve a better SEE performance with a reduced EH requirement. Note that the same SEE performance can be achieved with different available transmit power as in Figure  5 provided the SEE maximization problem is feasible with the same rest of the constraints. Hence, increasing the transmit power consumption cannot provide a better SEE. However, it can achieve a higher secrecy rate.
	
	  \begin{center}
		\begin{figure}[t]
			\centering\includegraphics[width=\linewidth]{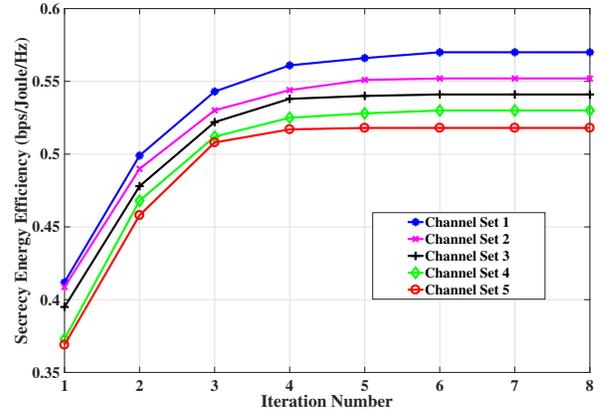}
			\caption{Convergence of the proposed algorithm with norm-error bounded imperfect CSI assumption in Algorithm 3 for different sets of channels.}
		\end{figure}
	\end{center}
	\begin{center}
		\begin{figure}[t!]
			\centering\includegraphics[width=\linewidth]{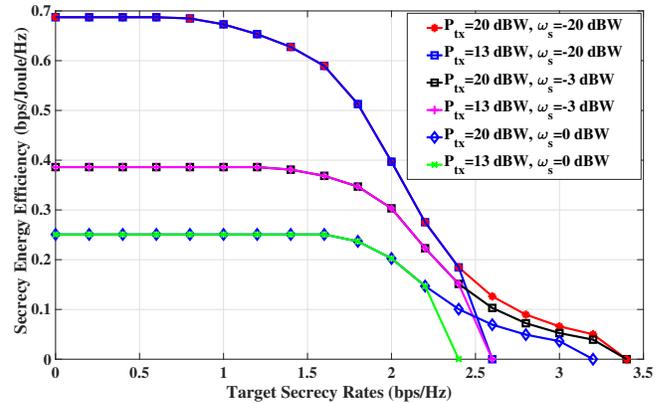}
			\caption{The achieved SEE with different target secrecy rates, transmit power constraints and EH requirements under the perfect CSI assumption.}
		\end{figure}
	\end{center}
	\vspace{-4em}
	
	We also perform performance comparisons between the proposed schemes and similar schemes available in the literature. Figure  6 provides comparisons of the achievable SEE of three different transmission schemes under the perfect CSI assumption: SEE maximization, power minimization and secrecy rate maximization. In these simulation results, the maximum available transmit power and EH requirement are assumed to be 20 dBW and $-20$ dBW, respectively. As expected, the proposed scheme for SEE maximization outperforms the other two schemes in terms of achieved SEE. Furthermore, the achievable SEE remains constant in the secrecy rate maximization scheme with different target secrecy rate values within its feasible region. This trend is due to the fact that the secrecy rate maximization scheme always achieves a better constant secrecy rate than the target secrecy rate in all cases with a feasible target secrecy rate. In other words, the target secrecy rate constraint becomes inactive with the available transmit power, which is more than the required to achieve the target secrecy rate. Furthermore, zero SEE means that it is not feasible to achieve the target secrecy rate with the available transmit power.
	
			\begin{center}
		\begin{figure}[t]
			\centering\includegraphics[width=\linewidth]{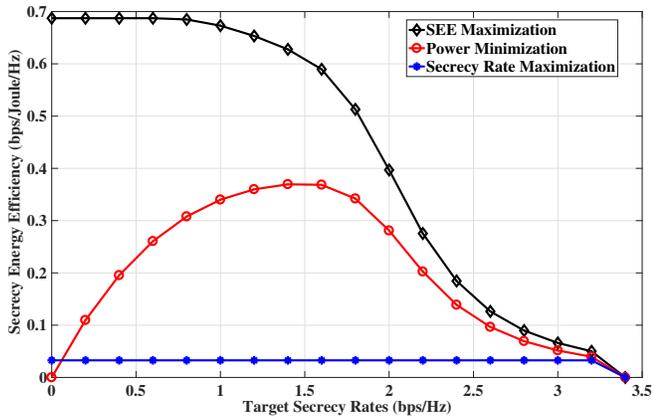}
			\caption{The achieved SEE for different transmission schemes under the assumption of perfect CSI: SEE maximization, power minimization and secrecy rate maximization.}
		\end{figure}
	\end{center}
	\begin{center}
		\begin{figure}[t]
			\centering\includegraphics[width=3.6 in]{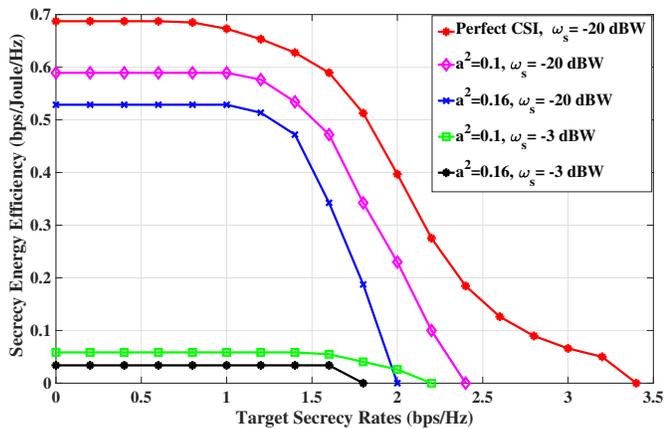}
			\caption{The achieved SEE with different target secrecy rates, channel covariances $a^{2}$ and EH requirements with the assumption of statistical CSI of ER channel.}
		\end{figure}
	\end{center}
	\vspace{-4em}

	We then evaluate the performance of the proposed scheme with statistical CSI on the ER's channel. Figure~7 shows the achieved SEE with different target secrecy rates and EH requirements. The ER's channel covariance matrix is modelled as $\mathbf{G}_{e}=a^{2}\mathbf{I}$. The parameters $\alpha, \beta$, and $\gamma$ define the outage probabilities for SEE, secrecy rate and EH requirement, respectively, and are assumed to be the same with all constraints, i.e., $\alpha=\beta=\gamma=0.1$. The parameter $a^{2}$ is introduced to control the ER's channel variance. Furthermore, we also evaluate the performance of the scheme with the perfect CSI to draw a performance comparison. As shown in Figure  7, the scheme with perfect CSI achieves the best performance and the optimal achievable SEE decreases as the target secrecy rate increases for all sets of the assumptions. Note that with a higher EH requirement and $a^{2}$, the proposed scheme achieves less SEE due to more required transmit power to meet the EH constraints and deal with the channel variance.

	The achieved SEE of different schemes is evaluated under the assumption of statistical CSI on the ER's channel. Figure 8 presents the achievable SEE of three schemes: SEE maximization, power minimization and secrecy rate maximization. Here, it is assumed that the maximum available transmit power and EH requirement are set to be $20$ dBW and $-20$ dBW, respectively. As expected, the proposed SEE maximization scheme outperforms the other two schemes in terms of the achieved SEE. Furthermore, the SEE performance obtained with the secrecy rate maximization scheme is not affected by different target secrecy rates in its feasible secrecy rate region. The reason is that in secrecy rate maximization scheme, all the available transmit power is consumed to achieve the maximum secrecy rate. Thus, the achievable SEE would remain a constant with a given available transmit power. 
			\begin{center}
		\begin{figure}[t!]
			\centering\includegraphics[width=\linewidth]{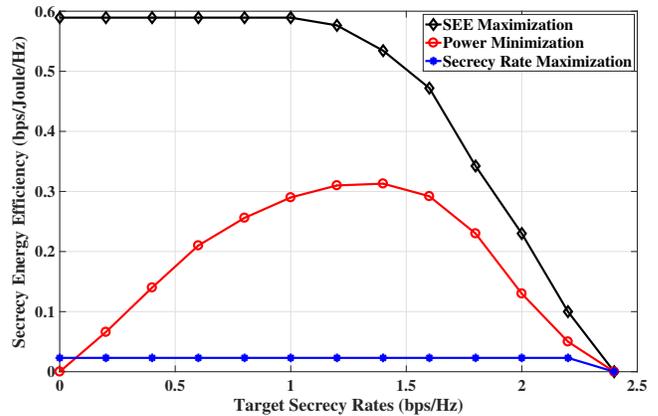}
			\caption{The achieved SEE with different schemes under the assumption of statistical CSI of ER channel: SEE maximization, power minimization and secrecy rate maximization.}
		\end{figure}
	\end{center}
		\begin{center}
		\begin{figure}[t]
			\centering\includegraphics[width=\linewidth]{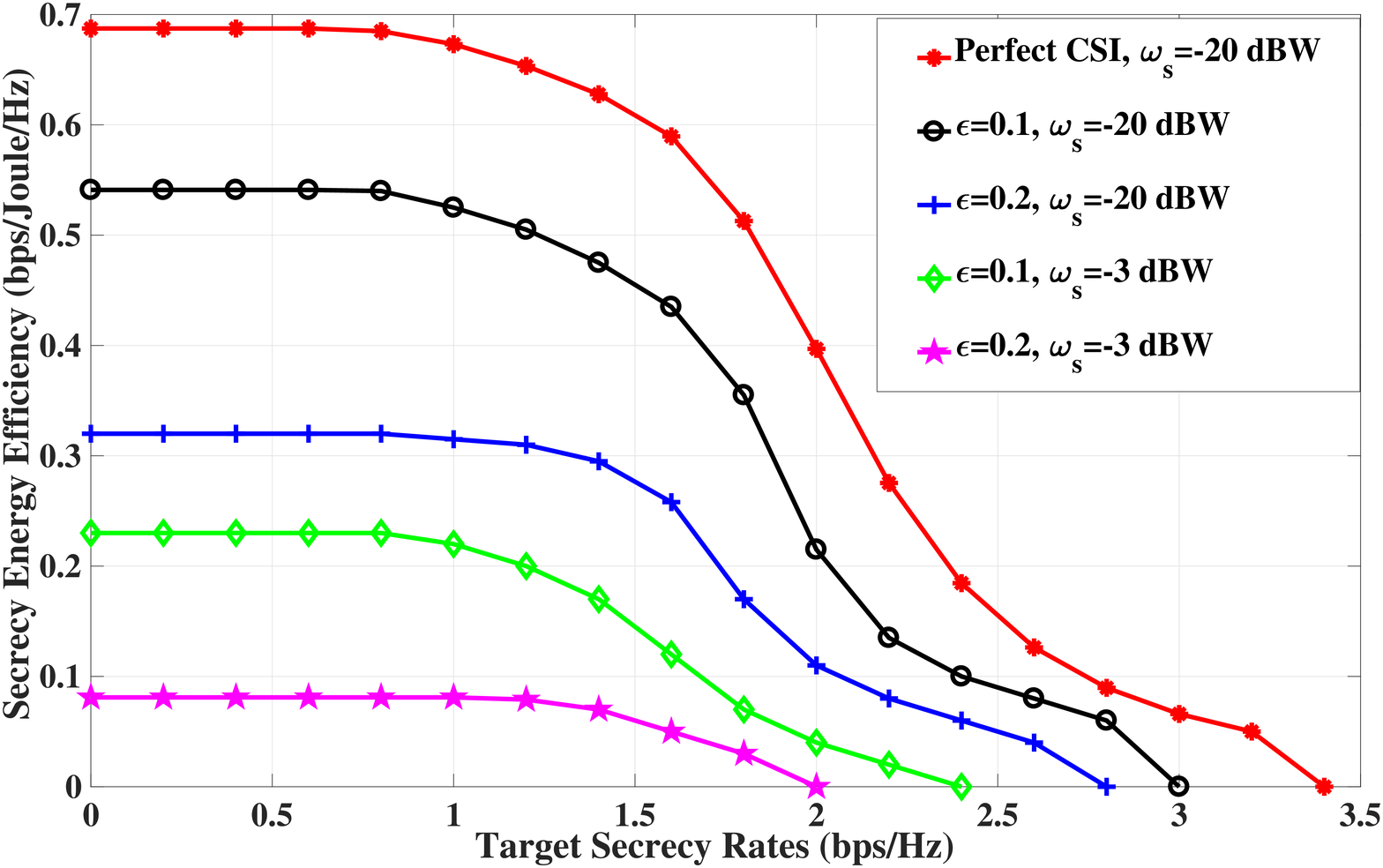}
			\caption{The achieved SEE with different target secrecy rates, norm-error bounds and EH requirements under the assumption of imperfect CSI.}
		\end{figure}
	\end{center}
	\vspace{-4em}
	
	Next we evaluate the performance of the proposed scheme under the assumption of the norm-bounded channel uncertainties in Section V. Figure 9 illustrates the achieved SEE with different target secrecy rates and EH requirements. The channel bounds of the uncertainties are assumed to be the same, i.e., $\epsilon_{s}=\epsilon_{p}=\epsilon_{e}=\epsilon$. The maximum available transmit power is set to be 20 dBW. As seen in Figure 9, the optimal SEE decreases as the target rate increases. The scheme with perfect CSI can be seen as a special case of the robust SEE maximization (with a zero error bound on the channel uncertainties, i.e., $\epsilon=0$). Similar to the previous results, zero SEE refers to an infeasible problem with a given target secrecy rate constraint. Furthermore, the optimal SEE has a better performance with a smaller EH requirement. Moreover, a better SEE can be realized for a feasible problem with a given target secrecy rate constraint and with a smaller error bound $\epsilon$. This is due to the fact that the SU-Tx requires more transmit power to deal with a larger channel uncertainty bound and to meet all set of constraints.
		
	\begin{center}
		\begin{figure}[t!]
			\centering\includegraphics[width=\linewidth]{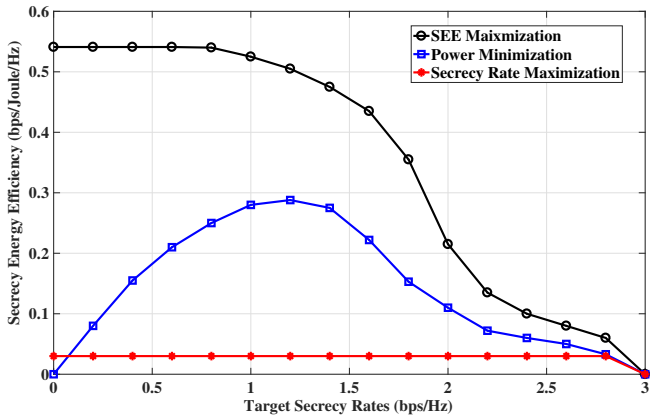}
			\caption{The achieved SEE with different schemes under the assumption imperfect CSI: SEE maximization, power minimization and secrecy rate maximization.}
		\end{figure}
	\end{center}
		\begin{center}
		\begin{figure}[t!]
			\centering\includegraphics[width=\linewidth]{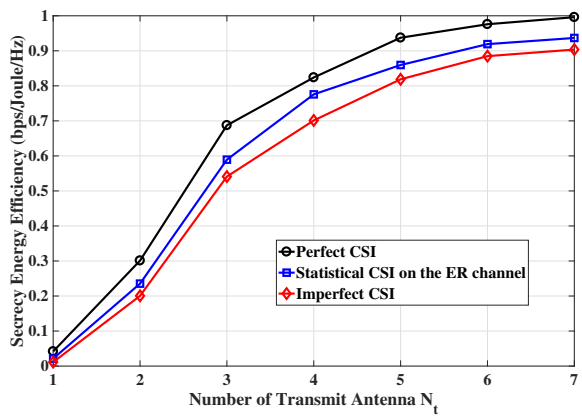}
			\caption{The achieved SEE with different number of transmit antennas with perfect CSI, statistical CSI on the ER channel and imperfect CSI assumptions. }
		\end{figure}
	\end{center}
	\vspace{-3em}
	
    We then provide the performance comparison of the achieved SEE with different transmission schemes under norm-error bounded channel uncertainties in Figure  10. Here, three different schemes are considered: SEE maximization, secrecy rate maximization and power minimization. Furthermore, the maximum available transmit power and the EH requirement are set to be 20 dBW and $-20$ dBW, respectively. As seen, our proposed SEE maximization scheme outperforms the other two schemes in terms of achieved SEE. Furthermore, the achievable SEE is not affected by the target secrecy rates and remains constant. This is due to the fact that all available transmit power is used to achieve the maximum secrecy rate in the secrecy maximization scheme, which is similar to that of the previous simulation results of different schemes under perfect and statistical CSI assumptions.
    	\begin{center}
    	\begin{figure}[t]
    		\centering\includegraphics[width=\linewidth]{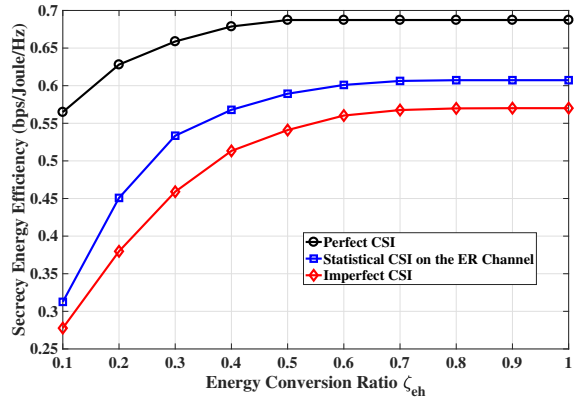}
    		\caption{The achieved SEE with different energy conversion ratio $\zeta_{eh}$ with perfect CSI, statistical CSI on the ER channel and imperfect CSI assumptions. }
    	\end{figure}
    \end{center}
	\vspace{-2.5em}

	The achieved SEE with different number of transmit antennas for all CSI assumptions is presented in Figure  11. The system parameters for all the cases are set to be as follows: target secrecy rate $R_{d} = 0.5$ bps/Hz, the transmit power consumption $P_{\textrm{tx}}=20$ dBW and the EH requirement $\omega_{s}=-20$ dBW. Furthermore, the ER's channel variance for the assumption of statistical CSI on the ER's channel is set to be $a^{2}=0.1$ and the channel uncertainty bound is assumed to be $\epsilon=0.1$ for the scheme with imperfect CSI, respectively. As shown in this figure, larger number of antennas at the SU-Tx provides better performance in terms of the SEE performance. This is due to the fact that larger number of transmit antennas brings more degree of freedom, which results in higher achievable secrecy rate without any additional transmit power consumption.
	
	Finally, we provide the SEE performance with different energy conversion ratio for all three CSI assumptions in Figure  12. Here, the target secrecy rate $R_{d}$, the transmit power consumption $P_{\textrm{tx}}$ and the EH requirement $\omega_{s}$ are set to be 0.5 bps/Hz, 20 dBW and -20 dBW, respectively. Furthermore, the ER's channel variance for the scheme with statistical CSI on the ER channel and the channel uncertainty bound for the scheme with imperfect CSI are assumed to be $a^{2}=0.1$ and $\epsilon=0.1$, respectively. As seen in Figure  12, the achieved SEE increases as the energy conversion ratio $\zeta_{eh}$ increases for all three CSI assumptions. The reason is that the SU-Tx requires more power to meet the EH requirement at the ER with a small energy conversion ratio $\zeta_{eh}$ (i.e., $\zeta_{eh}=0.1$).

	\section{Conclusions}
	\label{sec:conclusion}
	
	In this paper, we considered the SEE maximization problem for an underlay MISO CR network in the presence of energy harvesting. First, we proposed a transmit beamforming design to meet the required secrecy rate at the SU-Rx while satisfying the interference leakage constraint on the PU-Rx and the EH requirement on the ER. The original problem was not convex due to the non-linear fractional objective function. To overcome this non-convexity issue, we reformulated the original problem into a convex one by exploiting SDR, non-linear fractional and DC programming. We then extended the beamforming design to address the problems of imperfect CSI at the transmitter. In particular, two robust designs were proposed with the assumptions of statistical CSI on the ER's channel and error bounded channel uncertainties, respectively. For the case of statistical CSI assumption,  we first expressed the outage probability constraints in a closed-form expression. By exploiting SDR, non-linear fractional and DC programming, the original problem was efficiently solved through an iterative approach. Next, we formulated the robust design with bounded channel uncertainties as a series of SDP by exploiting the same techniques as in the previous robust design. Then, we used the S-procedure to convert this problem into a convex one. Simulation results were provided (1) to validate the convergence of the proposed algorithms, (2) to show the impacts of different parameters on the achieved SEE, including the maximum available transmit power, the EH requirement, the number of transmit antennas and the energy conversion ratio,  and (3) to demonstrate the superior performance of the proposed SEE maximization scheme over power minimization and secrecy rate maximization schemes available in the literature.

	\begin{appendices}
		\section{Proof of Theorem \ref{proof:nonlinear}}\label{proof_of_theorem}
		Let $\mathbf{Q}^{*}_{s}$ be the optimal solution of problem (\ref{eq:SEE_max_relax}), and $S$ be the set including all feasible $\mathbf{Q}_{s}$ under the constraints defined in (\ref{eq:SEE_Max_rec1})-(\ref{eq:SEE_Max_rec3}). Then, we have 
		\begin{align}
		\lambda^{*}=\frac{R_{s}(\mathbf{Q}^{*}_{s})}{P_{t}(\mathbf{Q}^{*}_{s})}\geq \frac{R_{s}(\mathbf{Q}_{s})}{P_{t}(\mathbf{Q}_{s})}, \textrm{for all}~ \mathbf{Q}_{s} \in S.
		\end{align}
		Hence,
		\begin{align}
		&R_{s}(\mathbf{Q}_{s})-\lambda^{*}P_{t}(\mathbf{Q}_{s})\leq 0, \textrm{for all}~ \mathbf{Q}_{s} \in S, \label{prooft1}\\
		&R_{s}(\mathbf{Q}^{*}_{s})-\lambda^{*}P_{t}(\mathbf{Q}^{*}_{s})=0.\label{prooft2}
		\end{align}
		From the inequality in (\ref{prooft1}) we have $\max\limits_{\mathbf{Q}_{s}} \{R_{s}(\mathbf{Q}_{s})-\lambda^{*}P_{t}(\mathbf{Q}_{s})\}=0$, and from (\ref{prooft2}) it is obvious that the maximum value is achieved with $\mathbf{Q}^{*}_{s}$. The sufficient condition has been satisfied, which completes the first part of the proof.
		
		Next, we consider the second part of the proof with the necessary condition. Let $\mathbf{Q}^{*}_{s}$ be the solution for $\max\limits_{\mathbf{Q}_{s}} \{R_{s}(\mathbf{Q}_{s})-\lambda^{*}P_{t}(\mathbf{Q}_{s})\}=0$ and $R_{s}(\mathbf{Q}^{*}_{s})-\lambda^{*}P_{t}(\mathbf{Q}^{*}_{s})=0.$ This implies the following:
		\begin{align}
		&R_{s}(\mathbf{Q}_{s})-\lambda^{*}P_{t}(\mathbf{Q}_{s})\leq 0, \textrm{for all}~ \mathbf{Q}_{s} \in S, \label{prooft3}\\
		&R_{s}(\mathbf{Q}^{*}_{s})-\lambda^{*}P_{t}(\mathbf{Q}^{*}_{s})=0.
		\end{align}
		From (\ref{prooft3}), we have $\lambda^{*}\geq \frac{R_{s}(\mathbf{Q}_{s})}{P_{t}(\mathbf{Q}_{s})}$ for all $\mathbf{Q}_{s} \in S$, and $\lambda^{*}$ is the maximum objective value of the problem defined in (\ref{eq:SEE_max_relax}). This completes the proof of Theorem \ref{proof:nonlinear}.\hfill $\blacksquare$
		\section{Proof of Proposition \ref{proposition:rank_proof1}}\label{proof_of_proposition1}
		First, we consider the following Lagrangian function of the optimization problem defined in (\ref{eq:SEE_max_relaxed}):
		\begin{align}
		&\mathcal{L}(\mathbf{Q}_{s},\mathbf{Z},\mu,\iota,\kappa,\varpi\!)\!=\!\!-\{\log_{2}(1\!+\!\frac{1}{\sigma_{su}^{2}}\mathbf{h}_{s}^{H}\mathbf{Q}_{s}\mathbf{h}_{s})\!-\!f(\mathbf{Q}_{s},\mathbf{Q}_{s}^{k})\nonumber\\
		&-\lambda[ \textrm{tr}(\mathbf{Q}_{s})+P_{c}]\}- \textrm{tr}(\mathbf{Z}\mathbf{Q}_{s})\!+\!\mu(\textrm{tr}(\mathbf{Q}_{s})- P_{\textrm{tx}})-\iota[\log_{2}(1\nonumber\\
		&+\frac{1}{\sigma_{su}^{2}}\mathbf{h}_{s}^{H}\mathbf{Q}_{s}\mathbf{h}_{s})-\log_{2}(1+\frac{1}{\sigma_{er}^{2}}\mathbf{h}_{e}^{H}\mathbf{Q}_{s}\mathbf{h}_{e})-R_{d}]- \kappa[\zeta_{eh}\nonumber\\
		&\mathbf{h}_{e}\mathbf{Q}_{s}\mathbf{h}_{e} - \omega_{s}]+\varpi(\mathbf{h}_{p}^{H}\mathbf{Q}_{s}\mathbf{h}_{p}-P_{f}),	
		\end{align}
		where $\mathbf{Q}_{s} \in \mathbb{H}^{N_{t}}_{+}$, $\mathbf{Z}\in \mathbb{H}^{N_{t}}_{+}$, $\mu\in\mathbb{R}_{+}$, $\iota\in\mathbb{R}_{+}$, $\kappa\in\mathbb{R}_{+}$ and $\varpi\in\mathbb{R}_{+}$ are the Lagrangian multipliers associated with the constraints in the problem defined in (\ref{eq:SEE_max_relaxed}). Then, we derive the corresponding  Karush-Kuhn-Tucker (KKT) conditions as follows~\cite{boyd2004convex}:
		\begin{align}
		&\frac{\partial\mathcal{L}}{\partial\mathbf{Q}_{s}}=-(\iota+1)\bigg[\frac{\frac{1}{\sigma_{su}^{2}}\mathbf{h}_{s}\mathbf{h}_{s}^{H}}{(1+\frac{1}{\sigma_{su}^{2}}\mathbf{h}_{s}^{H}\mathbf{Q}_{s}\mathbf{h}_{s})\ln2}\bigg] +(\lambda+\mu)\mathbf{I}+\nonumber\\
		&\iota\bigg[
		\frac{\frac{1}{\sigma_{er}^{2}}\mathbf{h}_{e}\mathbf{h}^{H}_{e}}{(1+\frac{1}{\sigma_{er}^{2}}\mathbf{h}_{e}^{H}\mathbf{Q}_{s}\mathbf{h}_{e})\ln2}\bigg]-\bigg[
		\frac{\frac{1}{\sigma_{er}^{2}}\mathbf{h}_{e}\mathbf{h}^{H}_{e}}{(1+\frac{1}{\sigma_{er}^{2}}\mathbf{h}_{e}^{H}\mathbf{Q}_{s}^{k}\mathbf{h}_{e})\ln2}\bigg]-\nonumber\\&\kappa\zeta_{eh}\mathbf{h}_{e}\mathbf{h}^{H}_{e}-\varpi\mathbf{h}_{p}\mathbf{h}^{H}_{p}-\mathbf{Z}=\mathbf{0},\\
		&\mathbf{Z}\mathbf{Q}_{s}=\mathbf{0}, \mathbf{Z}\succeq \mathbf{0}.
		\end{align}
		Based on these KKT conditions, the following equality holds:
		\begin{align} 
		&\!-\!(\iota\!+\!1)\bigg[\frac{\frac{1}{\sigma_{su}^{2}}\mathbf{h}_{s}\mathbf{h}_{s}^{H}}{(1\!+\!\frac{1}{\sigma_{su}^{2}}\mathbf{h}_{s}^{H}\mathbf{Q}_{s}\mathbf{h}_{s})\ln2}\bigg] \!+\!(\lambda\!+\!\mu)\mathbf{I}\!-\!\varpi\mathbf{h}_{p}\mathbf{h}^{H}_{p}\nonumber\\
		&\!+\!\iota\bigg[
		\frac{\frac{1}{\sigma_{er}^{2}}\mathbf{h}_{e}\mathbf{h}^{H}_{e}}{(1\!+\!\frac{1}{\sigma_{er}^{2}}\mathbf{h}_{e}^{H}\mathbf{Q}_{s}\mathbf{h}_{e})\ln2}\bigg]\!-\!\bigg[
		\frac{\frac{1}{\sigma_{er}^{2}}\mathbf{h}_{e}\mathbf{h}^{H}_{e}}{(1\!+\!\frac{1}{\sigma_{er}^{2}}\mathbf{h}_{e}^{H}\mathbf{Q}_{s}^{k}\mathbf{h}_{e})\ln2}\bigg]\nonumber\\&-\kappa\zeta_{eh}\mathbf{h}_{e}\mathbf{h}^{H}_{e}=\mathbf{Z},\label{rank_one_eq1}\\
		\Rightarrow & \bigg\{-(\iota+1)\bigg[\frac{\frac{1}{\sigma_{su}^{2}}\mathbf{h}_{s}\mathbf{h}_{s}^{H}}{(1+\frac{1}{\sigma_{su}^{2}}\mathbf{h}_{s}^{H}\mathbf{Q}_{s}\mathbf{h}_{s})\ln2}\bigg] +(\lambda+\mu)\mathbf{I}+\nonumber\\
		&\iota\bigg[
		\frac{\frac{1}{\sigma_{er}^{2}}\mathbf{h}_{e}\mathbf{h}^{H}_{e}}{(1+\frac{1}{\sigma_{er}^{2}}\mathbf{h}_{e}^{H}\mathbf{Q}_{s}\mathbf{h}_{e})\ln2}\bigg]-\bigg[
		\frac{\frac{1}{\sigma_{er}^{2}}\mathbf{h}_{e}\mathbf{h}^{H}_{e}}{(1+\frac{1}{\sigma_{er}^{2}}\mathbf{h}_{e}^{H}\mathbf{Q}_{s}^{k}\mathbf{h}_{e})\ln2}\bigg]\nonumber\\
		&-\kappa\zeta_{eh}\mathbf{h}_{e}\mathbf{h}^{H}_{e}-\varpi\mathbf{h}_{p}\mathbf{h}^{H}_{p} \bigg\}\mathbf{Q}_{s}=\mathbf{0},\\
		\Rightarrow&
		\bigg\{(\lambda+\mu)\mathbf{I}+\iota\bigg[
		\frac{\frac{1}{\sigma_{er}^{2}}\mathbf{h}_{e}\mathbf{h}^{H}_{e}}{(1+\frac{1}{\sigma_{er}^{2}}\mathbf{h}_{e}^{H}\mathbf{Q}_{s}\mathbf{h}_{e})\ln2}\bigg]-\nonumber\\
		&\bigg[
		\frac{\frac{1}{\sigma_{er}^{2}}\mathbf{h}_{e}\mathbf{h}^{H}_{e}}{(1+\frac{1}{\sigma_{er}^{2}}\mathbf{h}_{e}^{H}\mathbf{Q}_{s}^{k}\mathbf{h}_{e})\ln2}\bigg]-\kappa\zeta_{eh}\mathbf{h}_{e}\mathbf{h}^{H}_{e}\!-\!\varpi\mathbf{h}_{p}\mathbf{h}^{H}_{p} \bigg\}\mathbf{Q}_{s}\nonumber\\
		&
		=(\iota+1)\bigg[\frac{\frac{1}{\sigma_{su}^{2}}\mathbf{h}_{s}\mathbf{h}_{s}^{H}}{(1+\frac{1}{\sigma_{su}^{2}}\mathbf{h}_{s}^{H}\mathbf{Q}_{s}\mathbf{h}_{s})\ln2}\bigg]\mathbf{Q}_{s},\\
		\Rightarrow &
		\mathbf{Q}_{s}\!=\!\bigg\{(\iota\!+\!1)\bigg[\frac{\frac{1}{\sigma_{su}^{2}}\mathbf{h}_{s}\mathbf{h}_{s}^{H}}{(1\!+\!\frac{1}{\sigma_{su}^{2}}\mathbf{h}_{s}^{H}\mathbf{Q}_{s}\mathbf{h}_{s})\ln2}\bigg]\bigg\}\bigg\{(\lambda+\mu)\mathbf{I}+\nonumber\\
		&\iota\bigg[
		\frac{\frac{1}{\sigma_{er}^{2}}\mathbf{h}_{e}\mathbf{h}^{H}_{e}}{(1\!+\!\frac{1}{\sigma_{er}^{2}}\mathbf{h}_{e}^{H}\mathbf{Q}_{s}\mathbf{h}_{e})\ln2}\bigg]-\bigg[
		\frac{\frac{1}{\sigma_{er}^{2}}\mathbf{h}_{e}\mathbf{h}^{H}_{e}}{(1\!+\!\frac{1}{\sigma_{er}^{2}}\mathbf{h}_{e}^{H}\mathbf{Q}_{s}^{k}\mathbf{h}_{e})\ln2}\bigg]\nonumber\\&-\kappa\zeta_{eh}\mathbf{h}_{e}\mathbf{h}^{H}_{e}-\varpi\mathbf{h}_{p}\mathbf{h}^{H}_{p} \bigg\}^{-1}\mathbf{Q}_{s}.\label{rank_one_eq4}
		\end{align}
		Hence, the following rank relation holds:
		\begin{align}
		&\textrm{rank}(\mathbf{Q}_{s})=\textrm{rank}\bigg\{\bigg\{(\iota+1)\bigg[\frac{\frac{1}{\sigma_{su}^{2}}\mathbf{h}_{s}\mathbf{h}_{s}^{H}}{(1+\frac{1}{\sigma_{su}^{2}}\mathbf{h}_{s}^{H}\mathbf{Q}_{s}\mathbf{h}_{s})\ln2}\bigg]\bigg\}\nonumber\\
		&\bigg\{\iota\bigg[
		\frac{\frac{1}{\sigma_{er}^{2}}\mathbf{h}_{e}\mathbf{h}^{H}_{e}}{(1+\frac{1}{\sigma_{er}^{2}}\mathbf{h}_{e}^{H}\mathbf{Q}_{s}\mathbf{h}_{e})\ln2}\bigg]-\bigg[
		\frac{\frac{1}{\sigma_{er}^{2}}\mathbf{h}_{e}\mathbf{h}^{H}_{e}}{(1+\frac{1}{\sigma_{er}^{2}}\mathbf{h}_{e}^{H}\mathbf{Q}_{s}^{k}\mathbf{h}_{e})\ln2}\bigg]\nonumber\\&-\kappa\zeta_{eh}\mathbf{h}_{e}\mathbf{h}^{H}_{e}-\varpi\mathbf{h}_{p}\mathbf{h}^{H}_{p}+(\lambda+\mu)\mathbf{I} \bigg\}^{-1}\mathbf{Q}_{s}\bigg\}\nonumber\\
		&\leq\textrm{rank}\bigg[\frac{\frac{1}{\sigma_{su}^{2}}\mathbf{h}_{s}\mathbf{h}_{s}^{H}}{(1+\frac{1}{\sigma_{su}^{2}}\mathbf{h}_{s}^{H}\mathbf{Q}_{s}\mathbf{h}_{s})\ln2}\bigg]\leq 1.
		\end{align}
		By eliminating the trivial solution $\mathbf{Q}_{s}=\mathbf{0}$, we obtain $\textrm{rank}(\mathbf{Q}_{s})=1$, which completes the proof of Proposition 1.	\hfill $\blacksquare$	
		\section{Proof of Proposition \ref{proposition:rank_proof2}}\label{proof_of_proposition2}
		The proof of Proposition \ref{proposition:rank_proof2} is similar to that of Proposition 1 provided in Appendix B. First, we consider the following Lagrangian function of the problem defined in (\ref{SEE_max_st_dc}):
		\begin{align}
		&\mathcal{L}(\!\mathbf{Q}_{s},\mathbf{Z},\mu,\iota,\kappa,\varpi\!)\!=\!\!-\{\log_{2}(1\!\!+\!\!\frac{1}{\sigma_{su}^{2}}\mathbf{h}_{s}^{H}\mathbf{Q}_{s}\mathbf{h}_{s})\!-\!\!f_{s}(\mathbf{Q}_{s},\mathbf{Q}_{s}^{k})\nonumber\\
		&\!-\!\lambda[ \textrm{tr}(\mathbf{Q}_{s})+P_{c}]\}\!-\! \textrm{tr}(\mathbf{Z}\mathbf{Q}_{s})+\mu\textrm{tr}(\mathbf{Q}_{s}) - P_{\textrm{tx}})-\iota[1\!+\!\frac{1}{\sigma_{su}^{2}}\nonumber\\
		&\mathbf{h}_{s}^{H}\mathbf{Q}_{s}\mathbf{h}_{s}- 2^{R_{d}}+\frac{2^{R_{d}}}{\sigma_{er}^{2}}\textrm{tr}(\mathbf{G}_{e}\mathbf{Q}_{s})\ln\beta]+\kappa[\zeta_{eh}\textrm{tr}(\mathbf{G}_{e}\mathbf{Q}_{s})\nonumber\\
		&\ln(1-\gamma) + \omega_{s}]+\varpi(\mathbf{h}_{p}^{H}\mathbf{Q}_{s}\mathbf{h}_{p}-P_{f}),	
		\end{align}
		where $\mathbf{Q}_{s} \in \mathbb{H}^{N_{t}}_{+}$, $\mathbf{Z}\in \mathbb{H}^{N_{t}}_{+}$, $\mu\in\mathbb{R}_{+}$, $\iota\in\mathbb{R}_{+}$, $\kappa\in\mathbb{R}_{+}$ and $\varpi\in\mathbb{R}_{+}$ are the Lagrangian multipliers associated with problem (\ref{SEE_max_st_dc}). Then, we derive the corresponding  KKT conditions~\cite{boyd2004convex} as follows:
		\begin{align}
		\frac{\partial\mathcal{L}}{\partial\mathbf{Q}_{s}}=&-\bigg[\frac{\frac{1}{\sigma_{su}^{2}}\mathbf{h}_{s}\mathbf{h}_{s}^{H}}{(1+\frac{1}{\sigma_{su}^{2}}\mathbf{h}_{s}^{H}\mathbf{Q}_{s}\mathbf{h}_{s})\ln2}\bigg] +(\lambda+\mu)\mathbf{I}-\varpi\mathbf{h}_{p}\mathbf{h}^{H}_{p}\nonumber\\
		&+\bigg[
		\frac{-\frac{1}{\sigma_{er}^{2}}\mathbf{G}_{e}\ln\alpha}{(1-\frac{1}{\sigma_{er}^{2}}\textrm{tr}(\mathbf{G}_{e}\mathbf{Q}_{s}^{k})\ln\alpha)\ln2}\bigg]+\kappa\zeta_{eh}\textrm{tr}(\mathbf{G}_{e})\nonumber\\&-\iota\bigg[\frac{1}{\sigma_{su}^{2}}\mathbf{h}_{s}\mathbf{h}_{s}^{H}+
		\frac{2^{R_{d}}}{\sigma_{er}^{2}}\textrm{tr}(\mathbf{G}_{e})\ln\beta\bigg]-\mathbf{Z}=\mathbf{0}\nonumber\\
		\mathbf{Z}\mathbf{Q}_{s}=&\mathbf{0},~ \mathbf{Z}\succeq \mathbf{0}.
		\end{align}
		
		Following the same set of mathematical manipulations in (\ref{rank_one_eq1})-(\ref{rank_one_eq4}), the following rank relation holds:
		\begin{align}
		&\textrm{rank}(\mathbf{Q}_{s})=\textrm{rank}\bigg\{\bigg\{\bigg[\frac{\frac{1}{\sigma_{su}^{2}}\mathbf{h}_{s}\mathbf{h}_{s}^{H}}{(1+\frac{1}{\sigma_{su}^{2}}\mathbf{h}_{s}^{H}\mathbf{Q}_{s}\mathbf{h}_{s})\ln2}\bigg]\bigg\}\bigg\{(\lambda+\mu)\mathbf{I}\nonumber\\
		&+\bigg[
		\frac{-\frac{1}{\sigma_{er}^{2}}\mathbf{G}_{e}\ln\alpha}{(1-\frac{1}{\sigma_{er}^{2}}\textrm{tr}(\mathbf{G}_{e}\mathbf{Q}_{s}^{k})\ln\alpha)\ln2}\bigg]+\kappa\zeta_{eh}\textrm{tr}(\mathbf{G}_{e})-\varpi\mathbf{h}_{p}\mathbf{h}^{H}_{p}\nonumber\\
		&-\iota\bigg[\frac{1}{\sigma_{su}^{2}}\mathbf{h}_{s}\mathbf{h}_{s}^{H}+
		\frac{2^{R_{d}}}{\sigma_{er}^{2}}\textrm{tr}(\mathbf{G}_{e})\ln\beta\bigg] \bigg\}^{-1}\mathbf{Q}_{s}\bigg\}\nonumber\\
		&\leq\textrm{rank}\bigg[\frac{\frac{1}{\sigma_{su}^{2}}\mathbf{h}_{s}\mathbf{h}_{s}^{H}}{(1+\frac{1}{\sigma_{su}^{2}}\mathbf{h}_{s}^{H}\mathbf{Q}_{s}\mathbf{h}_{s})\ln2}\bigg]\leq 1.
		\end{align}
		By eliminating the trivial solution $\mathbf{Q}_{s}=\mathbf{0}$, we obtain $\textrm{rank}(\mathbf{Q}_{s})=1$, which completes the proof of Proposition 2.	\hfill $\blacksquare$
		\section{Proof of proposition \ref{proposition:rank_proof3}}	\label{proof_of_proposition3}
		First, suppose that we have solved the problem defined in (\ref{SEE_max_imperfecr_inner}) with the optimal value $\Lambda^{*}$. Now, the following power minimization problem is considered:
		\begin{align}\label{power_min}
		\min_{\mathbf{Q}_{s},\tau}&~\textrm{tr}(\mathbf{Q}_{s})\nonumber\\
		s.t.&~1+\frac{1}{\sigma_{su}^{2}}(\mathbf{\hat{h}}_{s}+\mathbf{\hat{e}}_{s})^{H}\mathbf{Q}_{s}(\mathbf{\hat{h}}_{s}+\mathbf{\hat{e}}_{s})\geq \tau \Lambda^{*},\nonumber\\
		&~\textrm{(\ref{SEE_max_imperfect_c1}),(\ref{SEE_max_imperfect_c2}),(\ref{eq:SEE_Max_imrec2}),(\ref{eq:SEE_Max_imrec4})},\nonumber \\
		&~\lambda(\textrm{tr}(\mathbf{Q}_{s})+P_{c})\leq \nu,~ \textrm{tr}(\mathbf{Q}_{s}) \leq P_{\textrm{tx}}, \mathbf{Q}_{s}\succeq \mathbf{0}.
		\end{align}
		By applying the S-procedure \cite{boyd1994linear}, the above problem can be transformed into the following SDP problem:
		\begin{align}\label{power_min_sp}
		\min_{\mathbf{Q}_{s},\tau}&~\textrm{tr}(\mathbf{Q}_{s})\nonumber\\
		s.t.&~\mathbf{T}_{1}(\mathbf{Q}_{s},\chi_{1})\succeq \mathbf{0}, \mathbf{T}_{2}(\mathbf{Q}_{s},\chi_{2})\succeq \mathbf{0}, \mathbf{T}_{3}(\mathbf{Q}_{s},\chi_{3})\succeq \mathbf{0},\nonumber\\
		&~\mathbf{T}_{4}(\mathbf{Q}_{s},\chi_{4})\succeq \mathbf{0}, \mathbf{T}_{5}(\mathbf{Q}_{s},\chi_{5})\succeq \mathbf{0},\nonumber\\ &~\lambda(\textrm{tr}(\mathbf{Q}_{s})+P_{c})\leq \nu, \textrm{tr}(\mathbf{Q}_{s}) \leq P_{\textrm{tx}}, \mathbf{Q}_{s}\succeq \mathbf{0},
		\end{align}
		where 
		\begin{align}
		&\mathbf{T}_{1}(\mathbf{Q}_{s},\chi_{1})=\nonumber\\
		&
		\left[
		\begin{matrix}
		\chi_{1}\mathbf{I}+\frac{1}{\sigma_{su}^{2}}\mathbf{Q}_{s} & \frac{1}{\sigma_{su}^{2}}\mathbf{Q}_{s}\mathbf{\hat{h}}_{s}\\
		\frac{1}{\sigma_{su}^{2}}\mathbf{\hat{h}}_{s}^{H}\mathbf{Q}_{s} & \frac{1}{\sigma_{su}^{2}}\mathbf{\hat{h}}_{s}^{H}\mathbf{Q}_{s}\mathbf{\hat{h}}_{s}+1-\tau\Lambda^{*}-\chi_{1}\epsilon_{s}^{2}
		\end{matrix}
		\right],\\
		&\mathbf{T}_{2}(\mathbf{Q}_{s},\chi_{2})=\nonumber\\
		&
		\left[
		\begin{matrix}
		\chi_{2}\mathbf{I}+\frac{1}{\sigma_{su}^{2}}\mathbf{Q}_{s} & \frac{1}{\sigma_{su}^{2}}\mathbf{Q}_{s}\mathbf{\hat{h}}_{s}\\
		\frac{1}{\sigma_{su}^{2}}\mathbf{\hat{h}}_{s}^{H}\mathbf{Q}_{s} & \frac{1}{\sigma_{su}^{2}}\mathbf{\hat{h}}_{s}^{H}\mathbf{Q}_{s}\mathbf{\hat{h}}_{s}+1-\tau2^{R_{d}}-\chi_{2}\epsilon_{s}^{2}
		\end{matrix}
		\right],\\
		&\mathbf{T}_{3}(\mathbf{Q}_{s},\chi_{3})=\nonumber\\
		&
		\left[
		\begin{matrix}
		\chi_{3}\mathbf{I}-\frac{1}{\sigma_{er}^{2}}\mathbf{Q}_{s} & -\frac{1}{\sigma_{er}^{2}}\mathbf{Q}_{s}\mathbf{\hat{h}}_{e}\\
		-\frac{1}{\sigma_{er}^{2}}\mathbf{\hat{h}}_{e}^{H}\mathbf{Q}_{s} & \tau-1-\frac{1}{\sigma_{er}^{2}}\mathbf{\hat{h}}_{e}^{H}\mathbf{Q}_{s}\mathbf{\hat{h}}_{e}-\chi_{3}\epsilon_{e}^{2}
		\end{matrix}
		\right],\\
		&\mathbf{T}_{4}(\mathbf{Q}_{s},\chi_{4})=\nonumber\\
		&
		\left[
		\begin{matrix}
		\chi_{4}\mathbf{I}+\zeta_{eh}\mathbf{Q}_{s} & \zeta_{eh}\mathbf{Q}_{s}\mathbf{\hat{h}}_{e}\\
		\zeta_{eh}\mathbf{\hat{h}}_{e}^{H}\mathbf{Q}_{s} & \zeta_{eh}\mathbf{\hat{h}}_{e}^{H}\mathbf{Q}_{s}\mathbf{\hat{h}}_{e}-\omega_{s}-\chi_{4}\epsilon_{e}^{2}
		\end{matrix}
		\right],\\
		&\mathbf{T}_{5}(\mathbf{Q}_{s},\chi_{5})\!=\!\left[
		\begin{matrix}
		\chi_{5}\mathbf{I}-\mathbf{Q}_{s} & -\mathbf{Q}_{s}\mathbf{\hat{h}}_{p}\\
		-\mathbf{\hat{h}}_{p}^{H}\mathbf{Q}_{s} & P_{f}-\mathbf{\hat{h}}_{p}^{H}\mathbf{Q}_{s}\mathbf{\hat{h}}_{p}-\chi_{5}\epsilon_{p}^{2}
		\end{matrix}
		\right].
		\end{align}
		Next, the Lagrangian function of the problem defined in (\ref{power_min_sp}) can be written as
		\begin{align}\label{lagelangri}
		\mathcal{L}(\mathbf{X})=&\textrm{tr}(\mathbf{Q}_{s})-\textrm{tr}(\mathbf{Q}_{s}\mathbf{Z})-\sum_{i=1}^{5}\textrm{tr}(\mathbf{T}_{i}(\mathbf{Q}_{s},\chi_{i})\mathbf{A}_{i})-\nonumber\\
		&\sum_{i=1}^{5}\chi_{i} m_{i},
		\end{align}
		where $\mathbf{X}$ denotes the set of all primal and dual variables: $\mathbf{X}=\{\mathbf{Q}, \mathbf{Z}, \tau, \mathbf{A}, \mathbf{\chi}, \mathbf{m}\}$
		with $\mathbf{A}=\{\mathbf{A}_{i}\}_{i=1}^{5}, \mathbf{\chi}=[\chi_{1},...,\chi_{5}]^{T}$ and $\mathbf{m}=[m_{1},...,m_{5}]^{T}$; $\mathbf{A}_{i} \in \mathbb{H}^{N_{t}+1}_{+}$, $\mathbf{Z}\in \mathbb{H}^{N_{t}}_{+}$ and $m_{i}\in\mathbb{R}_{+}$ are the Lagrangian multipliers associated with the problem in (\ref{power_min_sp}). For simple expressions, we rewrite $\mathbf{T}_{i}(\mathbf{Q}_{s},\chi_{i})$ as
		\begin{align}
		&\mathbf{T}_{1}(\mathbf{Q}_{s},\chi_{1})=\mathbf{S}_{1}(\mathbf{Q}_{s},\chi_{1})+\frac{1}{\sigma_{su}^{2}}\mathbf{V}_{s}^{H}\mathbf{Q}_{s}\mathbf{V}_{s},\label{bianhuan_1}
		\\
		&\mathbf{T}_{2}(\mathbf{Q}_{s},\chi_{2})=\mathbf{S}_{2}(\mathbf{Q}_{s},\chi_{2})+\frac{1}{\sigma_{su}^{2}}\mathbf{V}_{s}^{H}\mathbf{Q}_{s}\mathbf{V}_{s},\\
		&\mathbf{T}_{3}(\mathbf{Q}_{s},\chi_{3})=\mathbf{S}_{3}(\mathbf{Q}_{s},\chi_{3})-\frac{1}{\sigma_{er}^{2}}\mathbf{V}_{e}^{H}\mathbf{Q}_{s}\mathbf{V}_{e},\\
		&\mathbf{T}_{4}(\mathbf{Q}_{s},\chi_{4})=\mathbf{S}_{4}(\mathbf{Q}_{s},\chi_{4})+\zeta_{eh}\mathbf{V}_{e}^{H}\mathbf{Q}_{s}\mathbf{V}_{e},\\
		&\mathbf{T}_{5}(\mathbf{Q}_{s},\chi_{5})=\mathbf{S}_{5}(\mathbf{Q}_{s},\chi_{5})-\mathbf{V}_{p}^{H}\mathbf{Q}_{s}\mathbf{V}_{p},\label{bianhuan5}
		\end{align}
		where
		\begin{align}
		&\mathbf{S}_{1}(\mathbf{Q}_{s},\chi_{1})=\left[
		\begin{matrix}
		\chi_{1}\mathbf{I} & \mathbf{0}\\
		\mathbf{0} & 1\-\tau\Lambda^{*}-\chi_{1}\epsilon_{s}^{2}
		\end{matrix}
		\right],\nonumber\\
		&\mathbf{S}_{2}(\mathbf{Q}_{s},\chi_{2})=\left[
		\begin{matrix}
		\chi_{2}\mathbf{I} & \mathbf{0}\\
		\mathbf{0}& 1\!-\!\tau2^{R_{d}}\!-\!\chi_{2}\epsilon_{s}^{2}
		\end{matrix}
		\right],\nonumber\\
		&\mathbf{S}_{3}(\mathbf{Q}_{s},\chi_{3})=\left[
		\begin{matrix}
		\chi_{3}\mathbf{I} & \mathbf{0}\\
		\mathbf{0} & \tau\!-\!1\!-\!\chi_{3}\epsilon_{e}^{2}
		\end{matrix}
		\right],\nonumber\\
		&\mathbf{S}_{4}(\mathbf{Q}_{s},\chi_{4})=\left[
		\begin{matrix}
		\chi_{4}\mathbf{I} & \mathbf{0}\\
		\mathbf{0}& -\omega_{s}-\chi_{4}\epsilon_{e}^{2}
		\end{matrix}
		\right],\nonumber\\
		&\mathbf{S}_{5}(\mathbf{Q}_{s},\chi_{5})=\left[
		\begin{matrix}
		\chi_{5}\mathbf{I} & \mathbf{0}\\
		\mathbf{0} & P_{f}-\chi_{5}\epsilon_{p}^{2}
		\end{matrix}
		\right],\nonumber\\
		&\mathbf{V}_{s}=[\mathbf{I}_{N_{T}} ~\mathbf{\hat{h}_{s}}],  \mathbf{V}_{e}=[\mathbf{I}_{N_{T}} ~\mathbf{\hat{h}_{e}}], \mathbf{V}_{p}=[\mathbf{I}_{N_{T}} ~\mathbf{\hat{h}_{p}}].
		\end{align}
		Substituting (\ref{bianhuan_1})-(\ref{bianhuan5}) into (\ref{lagelangri}), we obtain a different expression for the Lagrangian function as follows:
		\begin{align}
		&\mathcal{L}(\mathbf{X})\!=\!\textrm{tr}(\mathbf{Q}_{s})-\textrm{tr}(\mathbf{Q}_{s}\mathbf{Z})-\textrm{tr}(\frac{1}{\sigma_{su}^{2}}\mathbf{Q}_{s}\mathbf{V}_{s}\mathbf{A}_{1}\mathbf{V}_{s}^{H})-\textrm{tr}(\frac{1}{\sigma_{su}^{2}}\nonumber\\
		&\mathbf{Q}_{s}\mathbf{V}_{s}\mathbf{A}_{2}\mathbf{V}_{s}^{H})\!+\!\textrm{tr}(\frac{1}{\sigma_{er}^{2}}\mathbf{Q}_{s}\mathbf{V}_{e}\mathbf{A}_{3}\mathbf{V}_{e}^{H})\!-\!\textrm{tr}(\zeta_{eh}\mathbf{Q}_{s}\mathbf{V}_{e}\mathbf{A}_{4}\mathbf{V}_{e}^{H})\nonumber\\
		&+\textrm{tr}(\mathbf{Q}_{s}\mathbf{V}_{p}\mathbf{A}_{5}\mathbf{V}_{p}^{H})+\varphi( \tau, \mathbf{A}, \mathbf{\chi}, \mathbf{m}),
		\end{align}
		where $\varphi( \tau, \mathbf{A}, \mathbf{\chi}, \mathbf{m})$ collects all the terms that are not related to $\mathbf{Q}_{s}$ and $\mathbf{Z}$, which are independent of this proof. Then, we define the KKT conditions~\cite{boyd2004convex} as:
		\begin{align}\label{kkt}
		\frac{\partial\mathcal{L}}{\partial\mathbf{Q}_{s}}=&\mathbf{I}-\mathbf{Z}-\frac{1}{\sigma_{su}^{2}}\mathbf{V}_{s}\mathbf{A}_{1}\mathbf{V}_{s}^{H}-\frac{1}{\sigma_{su}^{2}}\mathbf{V}_{s}\mathbf{A}_{2}\mathbf{V}_{s}^{H}+\nonumber\\
		&\frac{1}{\sigma_{er}^{2}}\mathbf{V}_{e}\mathbf{A}_{3}\mathbf{V}_{e}^{H}\!-\!\zeta_{eh}\mathbf{V}_{e}\mathbf{A}_{4}\mathbf{V}_{e}^{H}+\mathbf{V}_{p}\mathbf{A}_{5}\mathbf{V}_{p}^{H}=\mathbf{0},\\
		\mathbf{T}_{1}(\mathbf{Q}_{s}&,\chi_{1})\mathbf{A}_{1}=\mathbf{0},\label{second_step}\\
		\mathbf{Q}_{s}\mathbf{Z}=&\mathbf{0},  \mathbf{Z}\succeq \mathbf{0}, \mathbf{A}_{i}\succeq
		\mathbf{0}, i \in [1,2,3,4,5].
		\end{align}
		Premultiplying the two sides of (\ref{kkt}) and using the fact that $\mathbf{Q}_{s}\mathbf{Z}=\mathbf{0}$, we obtain
		\begin{align}
		&\mathbf{Q}_{s}\frac{1}{\sigma_{su}^{2}}\mathbf{V}_{s}\mathbf{A}_{1}\mathbf{V}_{s}^{H}\!=\!\mathbf{Q}_{s}\bigg( \mathbf{I}\!-\!\frac{1}{\sigma_{su}^{2}}\mathbf{V}_{s}\mathbf{A}_{2}\mathbf{V}_{s}^{H}\!+\!\frac{1}{\sigma_{er}^{2}}\mathbf{V}_{e}\mathbf{A}_{3}\mathbf{V}_{e}^{H}\nonumber\\
		&-\zeta_{eh}\mathbf{V}_{e}\mathbf{A}_{4}\mathbf{V}_{e}^{H}+\mathbf{V}_{p}\mathbf{A}_{5}\mathbf{V}_{p}^{H} \bigg).
		\end{align}
		The following rank relationship holds
		\begin{align}\label{diyige}
		&\textrm{rank}(\mathbf{Q}_{s})=\textrm{rank}\bigg(\mathbf{Q}_{s}\bigg(  \mathbf{I}-\frac{1}{\sigma_{su}^{2}}\mathbf{V}_{s}\mathbf{A}_{2}\mathbf{V}_{s}^{H}+\frac{1}{\sigma_{er}^{2}}\mathbf{V}_{e}\mathbf{A}_{3}\mathbf{V}_{e}^{H}\nonumber\\
		&\!-\!\zeta_{eh}\mathbf{V}_{e}\mathbf{A}_{4}\mathbf{V}_{e}^{H}\!+\!\mathbf{V}_{p}\mathbf{A}_{5}\mathbf{V}_{p}^{H} \bigg)\bigg)\!=\!\textrm{rank}(\mathbf{Q}_{s}\frac{1}{\sigma_{su}^{2}}\mathbf{V}_{s}\mathbf{A}_{1}\mathbf{V}_{s}^{H})\nonumber\\
		& \leq \min~ \{\textrm{rank}(\frac{1}{\sigma_{su}^{2}}\mathbf{V}_{s}\mathbf{A}_{1}\mathbf{V}_{s}^{H}), \textrm{rank}(\mathbf{Q}_{s})\}.
		\end{align}
		If it can be proven that $\textrm{rank}(\frac{1}{\sigma_{su}^{2}}\mathbf{V}_{s}\mathbf{A}_{1}\mathbf{V}_{s}^{H})=1$, then it is obvious that $\textrm{rank}(\mathbf{Q}_{s})\leq 1$. Hence, in the remaining part of this proof, we focus on the rank of $\frac{1}{\sigma_{su}^{2}}\mathbf{V}_{s}\mathbf{A}_{1}\mathbf{V}_{s}^{H}$. Substituting (\ref{bianhuan_1}) into (\ref{second_step}), we obtain
		\begin{align}
		\mathbf{S}_{1}(\mathbf{Q}_{s},\chi_{1})\mathbf{A}_{1}+\frac{1}{\sigma_{su}^{2}}\mathbf{V}_{s}^{H}\mathbf{Q}_{s}\mathbf{V}_{s}\mathbf{A}_{1}=\mathbf{0}.
		\end{align}
		Furthermore, it follows by post-multiplying by $\mathbf{V}_{s}^{H}$ that
		\begin{align}\label{kuaihaole}
		\mathbf{S}_{1}(\mathbf{Q}_{s},\chi_{1})\mathbf{A}_{1}\mathbf{V}_{s}^{H}+\frac{1}{\sigma_{su}^{2}}\mathbf{V}_{s}^{H}\mathbf{Q}_{s}\mathbf{V}_{s}\mathbf{A}_{1}\mathbf{V}_{s}^{H}=\mathbf{0}.
		\end{align}
		One can easily verify that
		\begin{align}
		&[\mathbf{I}_{N_{T}}~\mathbf{0}]\mathbf{S}_{1}(\mathbf{Q}_{s},\chi_{1})=\chi_{1}[\mathbf{I}_{N_{T}}~\mathbf{0}]=\chi_{1}(\mathbf{V}_{s}-[\mathbf{0}_{N_{T}}~\mathbf{\hat{h}_{s}}]),\nonumber\\
		&[\mathbf{I}_{N_{T}}~\mathbf{0}]\mathbf{V}_{s}^{H}=\mathbf{I}_{N_{T}}.
		\end{align}
		By pre-multiplying both sides of (\ref{kuaihaole}) by $[\mathbf{I}_{N_{T}}~\mathbf{0}]$, we obtain
		\begin{align}
		&\chi_{1}(\mathbf{V}_{s}-[\mathbf{0}_{N_{T}}~\mathbf{\hat{h}_{s}}])\mathbf{A}_{1}\mathbf{V}_{s}^{H}+\frac{1}{\sigma_{su}^{2}}\mathbf{Q}_{s}\mathbf{V}_{s}\mathbf{A}_{1}\mathbf{V}_{s}^{H}=\mathbf{0}\nonumber\\
		&\Leftrightarrow (\chi_{1}\mathbf{I}_{N_{T}}+\frac{1}{\sigma_{su}^{2}}\mathbf{Q}_{s})\mathbf{V}_{s}\mathbf{A}_{1}\mathbf{V}_{s}^{H}=\chi_{1}[\mathbf{0}_{N_{T}}~\mathbf{\hat{h}_{s}}]\mathbf{A}_{1}\mathbf{V}_{s}^{H}.
		\end{align}
		\begin{Lemma}(\!\!\cite{horn1990matrix}): If a block Hermitian matrix $$\mathbf{P}=\left[\begin{matrix} \mathbf{P}_{1} & \mathbf{P}_{2}\\
			\mathbf{P}_{3} & \mathbf{P}_{4}
			\end{matrix}\right]\succeq \mathbf{0},$$ then the main diagonal matrices $\mathbf{P}_{1}$ and $\mathbf{P}_{4}$ are always positive semidefinite matrices.
		\end{Lemma}
		
		Based on Lemma 4 and $\mathbf{T}_{1}(\mathbf{Q}_{s},\chi_{1})\succeq \mathbf{0}$, we can claim that $\chi_{1}\mathbf{I}_{N_{T}}+\frac{1}{\sigma_{su}^{2}}\mathbf{Q}_{s}$ is a positive definite matrix. Since multiplying both sides by a non-singular matrix does not change the rank of the original matrix, the following rank relationship holds:
		\begin{align}\label{dierge}
		&\textrm{rank}(\mathbf{V}_{s}\mathbf{A}_{1}\mathbf{V}_{s}^{H})=\textrm{rank}\bigg((\chi_{1}\mathbf{I}_{N_{T}}+\frac{1}{\sigma_{su}^{2}}\mathbf{Q}_{s})\mathbf{V}_{s}\mathbf{A}_{1}\mathbf{V}_{s}^{H}\bigg)\nonumber\\
		&=\textrm{rank}(\chi_{1}[\mathbf{0}_{N_{T}}~\mathbf{\hat{h}_{s}}]\mathbf{A}_{1}\mathbf{V}_{s}^{H})=\textrm{rank}([\mathbf{0}_{N_{T}}~\mathbf{\hat{h}_{s}}])\leq 1.
		\end{align}
		Combining (\ref{diyige}) and (\ref{dierge}), we have
		\begin{align}
		\textrm{rank}(\mathbf{Q}_{s})\leq\textrm{rank}(\mathbf{V}_{s}\mathbf{A}_{1}\mathbf{V}_{s}^{H})\leq 1.
		\end{align}
		Similar to the previous proofs, by eliminating the trivial solution $\mathbf{Q}_{s}=\mathbf{0}$, we obtain $\textrm{rank}(\mathbf{Q}_{s})=1$, which completes the proof of Proposition 3.	\hfill $\blacksquare$
	\end{appendices}
	\ifCLASSOPTIONcaptionsoff
	\newpage
	\fi

	\bibliographystyle{IEEEtran}
	
	\bibliography{referenceIEEE}

\end{document}